\documentclass[mathpazo]{jmlc}

\usepackage[T1]{fontenc}
\usepackage{microtype}
\usepackage{graphicx}
\usepackage{tikz}
\usepackage{booktabs}
\usepackage{mathtools}
\usepackage{algorithm}
\usepackage{algorithmic}
\usepackage[most]{tcolorbox}
\usepackage{caption}
\usepackage{multirow,multicol,array}
\usepackage{fvextra}
\usepackage{enumitem}
\usepackage{xcolor}
\colorlet{blue}{black}
\usepackage{placeins}
\usepackage[hidelinks]{hyperref}

\newcommand{\citep}{\cite}

\newcommand{\sysname}{ReasModel}

\definecolor{humanbubble}{RGB}{235,245,255}
\definecolor{aibubble}{RGB}{245,245,245}
\definecolor{loggray}{RGB}{230,230,230}
\definecolor{indexbg}{RGB}{248,248,248}
\definecolor{reaslab}{RGB}{112,48,160}
\definecolor{reaslight}{RGB}{246,243,255}
\newtcolorbox{interactionlog}[1][]{
  enhanced,
  arc=0pt, outer arc=0pt,
  colback=white, colframe=reaslab,
  boxrule=0.8pt,
  fonttitle=\bfseries\sffamily,
  coltitle=reaslab, colbacktitle=reaslight,
  title={#1},
  halign title=center, attach title to upper,
  after title={\vspace{4pt}\hrule\vspace{10pt}},
}
\fvset{breaklines=true}
\hypersetup{
  pdftitle={ReasModel: An Asynchronous Multi-Agent System for End-to-End Mathematical Modeling},
  pdfauthor={Can Ren, Libin Zheng, Yuhao Jiang, Xinyu Chen, Shumeng Xia, Zhiyi Guo, Luhong Guo, Zhewei Zhang, Jiang Hu, Zaiwen Wen}
}

\begin{document}

\title[ReasModel]{ReasModel: An Asynchronous Multi-Agent System for End-to-End Mathematical Modeling}
\author[C. Ren et al.]{
\mbox{Can Ren}\affil{1,*}, \mbox{Libin Zheng}\affil{2,*}, \mbox{Yuhao Jiang}\affil{3,*},
\mbox{Xinyu Chen}\affil{1}, \mbox{Shumeng Xia}\affil{4}, \mbox{Zhiyi Guo}\affil{1},
\mbox{Luhong Guo}\affil{5}, \mbox{Zhewei Zhang}\affil{6}, \mbox{Jiang Hu}\affil{7,\dagger},
and \mbox{Zaiwen Wen}\affil{8}}
\address{\setlength{\baselineskip}{10pt}
\affilnum{1} School of Mathematical Sciences, Peking University, Beijing, China.\\
\affilnum{2} Center for Intelligent Computing, Great Bay University, Dongguan, China.\\
\affilnum{3} School of Mathematics, Hunan University, Changsha, China.\\
\affilnum{4} College of Engineering, Peking University, Beijing, China.\\
\affilnum{5} Beijing-Dublin International College, Beijing University of Technology, Beijing, China.\\
\affilnum{6} College of Mathematics and Systems Science, Shandong University of Science and Technology, Qingdao, China.\\
\affilnum{7} Yau Mathematical Sciences Center, Tsinghua University, Beijing, China.\\
\affilnum{8} Beijing International Center for Mathematical Research, Peking University, Beijing, China.\endgraf\vspace{-10pt}}

\begin{abstract}
Mathematical modeling requires the integration of domain knowledge, mathematical
reasoning, data analysis, executable computation, and scientific communication into
a coherent solution. We present ReasModel, an end-to-end agent that
coordinates specialized modeler, coder, and writer roles through asynchronous
multi-type scheduling. Its \textcolor{blue}{3}-layer communication and grounding mechanism
combines coordinator-mediated information forwarding, indexed per-agent workspaces,
and artifact metadata to preserve modeling decisions and computational evidence
throughout the workflow. We further introduce MathModeling-Bench, a Chinese-language,
report-native benchmark comprising 46 authentic CUMCM problems, their original
statements and \textcolor{blue}{contest-provided data and supporting files, where supplied}, and 282 weighted problem-specific requirements.
The benchmark evaluates complete reports in terms of problem-specific fulfillment,
general modeling quality, and visual quality. \textcolor{blue}{ReasModel outperforms
the evaluated base-model and existing-agent baselines on both our newly constructed
MathModeling-Bench and the public MM-Bench, gaining 18.25 points over the strongest
baseline on MathModeling-Bench. On MM-Bench, it also reduces output-token consumption
by 23.7\% and runtime by 32.6\% compared with the existing LLM-MM-Agent under the
same backbone.}

\end{abstract}
\footnotetext[1]{Equal contribution.}
\footnotetext[2]{Corresponding author: Jiang Hu.}
\maketitle
\setcounter{footnote}{0}
\renewcommand{\thefootnote}{\arabic{footnote}}

\section{Introduction}
\label{sec:intro}

Mathematical modeling translates underspecified real-world situations into
mathematical representations that can be analyzed, computed, and acted upon.
It is important across scientific research and industrial practice in part
because most decision problems do not arrive pre-formalized: variables,
assumptions, objectives, and constraints must themselves be identified and
justified before any computation can begin. The modeler must determine which
information is relevant, introduce defensible assumptions, select an
appropriate model family, connect heterogeneous documents and data to a
coherent formulation, implement and solve the resulting problem, interpret
outputs in context, and revise the model when validation reveals a
mismatch. This makes modeling fundamentally iterative---a cycle of
formulation, computation, interpretation, and revision---rather than a
one-shot derivation~\citep{garfunkel2019gaimme}.

The ubiquity of modeling makes its automation a problem of both practical
value and scientific interest: it tests whether AI systems can act
competently on open-ended, real-world situations rather than well-posed
exercises. Recent advances in large language models (LLMs) have spurred
progress along this front, from translating natural-language descriptions
into executable mathematical
formulations~\citep{huang2025mamo,ramamonjison2023nl4opt} to producing
end-to-end solutions for competition-style modeling
problems~\citep{liu2025mmagent,qian2025modelingagent,jia2026scimind}.
Despite this progress, existing systems do not necessarily enforce
consistency among assumptions, formulations, executable code, numerical
outputs, and the final narrative. As a result, an early modeling or
implementation error can persist undetected into a fluent final report.

Tool use provides a mechanism for reducing this disconnect by exposing
intermediate computations, generated artifacts, and execution feedback to
later stages. In the ReAct pattern, an agent interleaves reasoning with tool
invocations, observes the results, and uses those observations to guide
subsequent steps~\citep{yao2023react}. This interaction turns external tools
from passive utilities into sources of evidence: a parser can reveal the
actual structure of an attachment, a data-analysis program can expose
missing values or unexpected distributions, and a solver can return
infeasibility certificates, numerical diagnostics, or candidate solutions
that challenge the agent's initial assumptions. Applied to mathematical
modeling, tool use therefore supports not only task execution---including
inspecting attachments, processing data, solving optimization or simulation
models, generating figures, and compiling reports---but also iterative
verification and correction. Intermediate outputs can be checked against
the formulation, reused by downstream stages, and traced back when the
final results appear inconsistent. By grounding later decisions in
inspectable artifacts rather than solely in the model's textual reasoning,
a tool-augmented workflow can make the connection between assumptions,
code, numerical evidence, and reported conclusions more explicit and
auditable.

However, a complete modeling workflow couples activities with different
durations and dependencies: model design, computation, visualization, and
writing can sometimes proceed concurrently where dependencies permit but
must also exchange intermediate decisions and artifacts. A strictly serial
pipeline prevents independent stages from overlapping; forwarding full
interaction histories, on the other hand, consumes context and risks mixing
discarded alternatives with final decisions. Existing agent-based systems
have demonstrated that tool-augmented pipelines can produce complete
modeling reports~\citep{liu2025mmagent,qian2025modelingagent}, yet their
\textcolor{blue}{computational cost, measured in token consumption, tool invocations, or
wall-clock time, can be substantial relative to the quality gains they}
deliver, motivating architectures that achieve comparable output quality
with leaner resource budgets. The challenge is therefore not merely one of
tool access but of coordination: how can specialized roles proceed
concurrently where dependencies permit and hand off compact, inspectable
artifacts while preserving the information needed for a coherent final
report?

Building an agent capable of executing such a workflow is only half the
problem; evaluating its output poses a second, equally demanding challenge.
Open-ended modeling admits multiple defensible formulations, so
exact-answer matching is too restrictive. At the same time, a fluent report
may omit a requested subproblem, constraint, or analysis; it may nominally
cover the task while remaining weak in problem understanding, modeling
correctness, validation, or sensitivity analysis; and it may communicate
results poorly through its figures, tables, or document organization. A
single holistic score conceals these distinct failure modes. Moreover,
requirement-based scoring can inadvertently encourage over-generation when
additional coverage carries little downside. Excessive content often
manifests in the rendered report as dense prose, repetitive presentation,
weak information hierarchy, and poor page composition. No single metric fully captures these effects. The visual dimension captures some practical consequences of over-generation, although it is not intended as a direct measure of conciseness. The challenge is therefore not merely one of scoring but of evaluation design: how can such reports be evaluated along complementary dimensions without collapsing distinct failure modes into a single score or inadvertently rewarding over-generation?

We address the systems and evaluation challenges jointly.
\textsc{ReasModel} is an asynchronous multi-agent framework in
which a coordinator schedules specialized modeler, coder, and writer agents
over a shared workspace. It combines asynchronous multi-type scheduling
with a \textcolor{blue}{3}-layer communication and grounding contract---coordinator-mediated
forwarding, per-agent subworkspaces with \texttt{README} indices, and sidecar
\texttt{.metadata} files---to coordinate model construction, executable
computation, visualization, and report writing. We pair the system with
\textsc{MathModeling-Bench}, a report-native benchmark of 46 authentic
China Undergraduate Mathematical Contest in Modeling (CUMCM) problems from
2011--2025. The benchmark retains the original problem statements and all
\textcolor{blue}{contest-provided data and supporting files, including spreadsheets,
CSV files, documents, images, and video where supplied,} and contains 282 weighted, separately scorable
problem-specific requirements. Its evaluation protocol combines
Problem-Specific Rubric Fulfillment (PSRF), computed from these
requirements, with general rubrics for Modeling Solution Quality (MSQ) and
Report Visual Quality (RVQ), applied to complete rendered reports.

To examine practical use beyond controlled benchmark evaluation, we deploy
the system through ReasLab,\footnote{ReasLab can be accessed at
\texttt{https://reaslab.io}.} where users can work
with original problem files, persistent artifacts, and compiled reports,
and release an open
ReasLab ReasModel Codex Bundle\footnote{The ReasLab ReasModel
Codex Bundle can be accessed at
\texttt{https://github.com/reaslab/MathModelingAgent-CodexBundle}.} for local
inspection and reuse. We
additionally report an \textcolor{blue}{8}-month production deployment involving 381
users and 763 projects; these statistics measure operational uptake rather
than task success and are presented separately from benchmark quality in
Section~\ref{sec:deployment}. \textcolor{blue}{2} industry-inspired case studies further examine sparse MIMO array design
and open-world encrypted-traffic recognition. Their synthetic proxy experiments
illustrate how physical approximation, finite-sample uncertainty, and
implementation fidelity affect modeling decisions.

Our contributions are threefold:
\begin{itemize}[leftmargin=*]
    \item We present \textsc{ReasModel}, an end-to-end mathematical
    modeling system that combines asynchronous multi-type scheduling with a
    \textcolor{blue}{3}-layer communication and grounding contract (coordinator-mediated
    forwarding, per-agent subworkspaces with \texttt{README} indices, and
    sidecar \texttt{.metadata} files) for coordinating model construction,
    executable computation, visualization, and report writing.

    \item We introduce \textsc{MathModeling-Bench}, a report-native
    benchmark comprising 46 authentic CUMCM problems with their original
    statements and \textcolor{blue}{contest-provided data and supporting files, where supplied}, 282 problem-specific requirements,
    and a multidimensional PSRF/MSQ/RVQ evaluation protocol. \textcolor{blue}{Across
    MathModeling-Bench and the public MM-Bench, \textsc{ReasModel} consistently
    outperforms the evaluated base-model and existing-agent baselines in overall
    evaluation scores. It also generates complete reports with fewer output tokens
    and shorter runtime than LLM-MM-Agent under matched backbones.}

    \item We study practical applicability beyond benchmark scores through
    \textcolor{blue}{2} industry-inspired modeling cases---sparse MIMO array design and
    open-world encrypted-traffic recognition---that expose the roles of
    physical approximation, finite-sample uncertainty, and implementation
    fidelity in end-to-end modeling. We complement these scoped synthetic
    proxy studies with an \textcolor{blue}{8}-month production deployment on ReasLab and
    release a public Codex bundle for inspection and reuse. The deployment
    records operational adoption rather than task success, and the case
    studies are not presented as full industrial validation.
\end{itemize}

\section{Related Work}\label{sec:related}

\subsection{From Mathematical Reasoning to Mathematical Formalization}

Mathematical reasoning concerns solving problems whose mathematical structure
is already largely specified. Chain-of-Thought prompting shows that explicit
intermediate reasoning can substantially improve multi-step problem solving
\citep{wei2022chainofthought}. Minerva demonstrates the value of continued
pretraining on technical corpora \citep{lewkowycz2022minerva}, while
MetaMath improves open-source models through large-scale synthetic
mathematical instruction data \citep{yu2024metamath}. More recently,
DeepSeekMath combines mathematical pretraining with reinforcement-learning
post-training to further strengthen mathematical reasoning
\citep{shao2024deepseekmath}.

Mathematical formalization addresses a different problem: constructing a
mathematical representation from natural-language descriptions. NL4Opt
studies natural-language-to-optimization formulation
\citep{ramamonjison2023nl4opt}, ORLM trains specialized models for
optimization modeling \citep{huang2024orlm}, and Mamo extends evaluation to
multiple formalisms such as differential equations, linear programs, and
mixed-integer programs \citep{huang2025mamo}. These works typically assume
that the target modeling paradigm is largely known in advance, whereas our
work considers broader mathematical modeling that additionally requires
method selection, computation, analysis, visualization, and report
generation.

\subsection{Agentic Workflows for Long-Horizon Tasks}

Long-horizon agentic workflows address tasks that cannot be completed through
a single model response and instead require multiple rounds of planning,
external actions, observation, and revision. Such systems typically
decompose a complex objective into intermediate steps and use tools or
persistent outputs to ground subsequent decisions. ReAct and Plan-and-Solve
demonstrate the value of task decomposition and iterative reasoning
\citep{yao2023react,wang2023planandsolve}, while Toolformer and Reflexion
show how external tools and feedback can be incorporated into this process
\citep{schick2023toolformer,shinn2023reflexion}. SWE-agent further
illustrates this paradigm in software engineering, where repository
navigation, file editing, code execution, and testing form an iterative
action--feedback loop \citep{yang2024sweagent}.

Recent systems extend these ideas to data analysis and scientific research.
Data Interpreter addresses complex data science workflows through
graph-structured planning and progressive execution
\citep{hong2025datainterpreter}. The AI Scientist integrates idea
generation, code implementation, experimentation, visualization, and paper
writing into an automated research process \citep{lu2024aiscientist}.
ReasFlow further targets reasoning-centric scientific discovery in applied
mathematics, combining literature synthesis, algorithm design, theorem
proving, numerical experiments, and manuscript preparation through
specialized agents \citep{he2026reasflow}. These systems demonstrate how
agents can coordinate extended workflows over code, results, figures, and
documents. Our work builds on this long-horizon perspective, but focuses
specifically on coordinating the heterogeneous and mutually dependent
artifacts produced during mathematical modeling.

\subsection{Agents for Mathematical Modeling}

Mathematical modeling agents aim to automate not only mathematical reasoning,
but the broader process of translating an underspecified real-world problem
into a defensible model and computational solution. Existing systems can be
roughly divided into \emph{formalism-constrained} agents, where the target
mathematical representation is largely known in advance, and
\emph{open-ended} agents, which must also determine the modeling approach
itself.

OptiMUS and OR-LLM-Agent belong primarily to the former category. They
combine mathematical formulation, solver-code generation, execution, and
iterative debugging for operations-research problems under predefined
optimization formalisms
\citep{ahmaditeshnizi2024optimus,zhang2025orllmagent}. More recent work
targets open-ended competition-style modeling. MM-Agent decomposes the
process into problem analysis, mathematical modeling, computational solving,
and report generation \citep{liu2025mmagent}. ModelingAgent coordinates
specialized roles and iterative refinement for interdisciplinary modeling
problems \citep{qian2025modelingagent}, while Sci-Mind introduces
\textcolor{blue}{experience-based retrieval} and adversarial verification to improve mathematical
coherence and computational feasibility \citep{jia2026scimind}.

Unlike prior systems that organize modeling primarily as a fixed sequence of
stages, \sysname{} uses asynchronous, dependency-aware coordination over
shared artifacts. This design allows formulations, code, results, figures,
and report sections to be produced and revised as their dependencies become
available.

\subsection{Agent Benchmarks and Evaluation}

The move from solver-verifiable optimization outputs to full
competition-style reports also changes what agent benchmarks must measure.
Recent benchmarks for LLM agents increasingly emphasize authentic task
contexts, persistent environments, and outcome-grounded evaluation
\citep{liu2024agentbench,jimenez2024swebench,chan2025mlebench}. In
software and machine learning engineering, benchmarks such as SWE-bench
and MLE-bench evaluate whether agents can resolve real issues or complete
repository-centered workflows under realistic constraints
\citep{jimenez2024swebench,chan2025mlebench}. Their use in evaluating
frontier systems such as DeepSeek-V4 and GPT-6 Astra further suggests
that SWE-bench and MLE-bench have become widely recognized tools for
assessing agent capabilities
\citep{deepseek2026v4report,openai2026gpt6astrasystemcard}.
Interactive web and computer-use benchmarks, by contrast, stress planning,
tool use, and error recovery in open-ended environments
\citep{zhou2024webarena,xie2024osworld}. Together, these benchmarks show
that realistic agent evaluation should preserve task structure and assess
success through inspectable artifacts and externally meaningful outcomes.

Open-ended mathematical modeling makes these demands even sharper, as
reflected in mathematical modeling benchmarks such as MM-Bench
\citep{liu2025mmagent}. MM-Bench evaluates agents on authentic
mathematical modeling contest problems. Sci-Mind further uses EngiBench
alongside MM-Bench to assess whether its framework can generalize beyond
mathematical modeling to broader open-ended modeling tasks, highlighting
EngiBench's role as a benchmark for broader engineering problem solving
\citep{jia2026scimind,zhou2026engibench}. Such tasks allow multiple
defensible formulations, yet still require problem-specific assumptions,
data use, computations, and final reporting. Accordingly, exact-match
evaluation is too restrictive for open-ended modeling, while
unconstrained holistic judgment can obscure whether a report actually
satisfies requested subproblems, constraints, analyses, and validation
requirements. Requirement-oriented scoring is also insufficient on its
own, since it can reward over-coverage without showing whether the
system truly understands the problem or produces a strong model and
report. Unlike benchmarks centered on structured outputs, our benchmark
evaluates complete rendered reports using complementary measures of task
fulfillment, modeling quality, and visual communication.

\section{The \sysname{} System}\label{sec:system}

\sysname{} is organized around a shared workspace rather than a fixed
pipeline. This section explains how that organization lets modeling,
computation, and writing proceed concurrently, and how the resulting
artifacts are handed off so that every claim in the report can be
traced to a computation.

\subsection{System Overview}\label{sec:overview}

\begin{figure}[!htbp]
\centering
\includegraphics[width=0.88\linewidth,trim=29bp 20bp 9bp 1bp,clip]{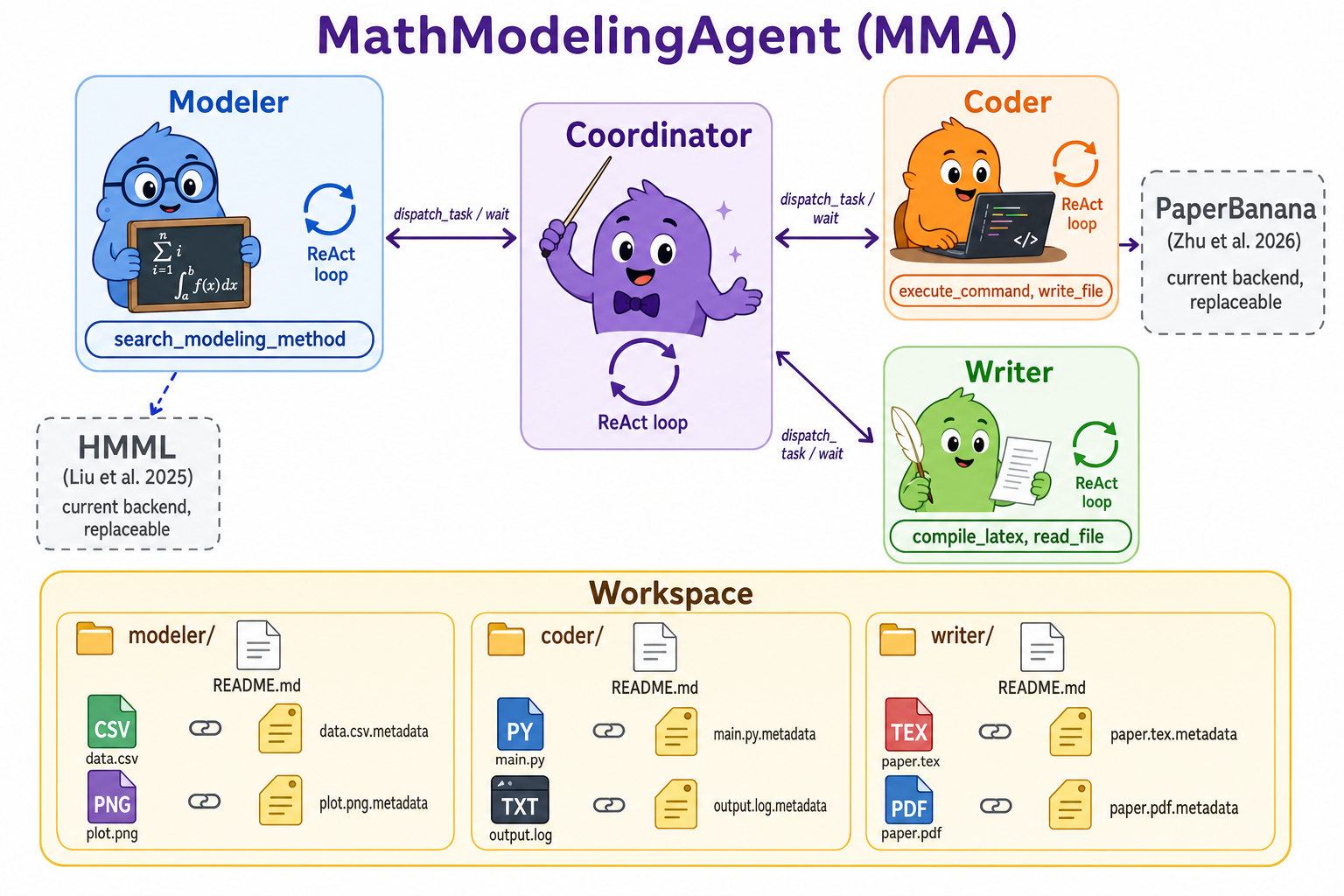}
\caption{Architecture of \sysname{}: a Coordinator orchestrates \textcolor{blue}{3}
specialized ReAct workers (Modeler, Coder, Writer) over a shared
workspace partitioned into per-agent subdirectories with
\texttt{README.md} indices and artifact / \texttt{.metadata} sidecar
pairs.}
\label{fig:system_overview}
\end{figure}

\sysname{} is a multi-agent framework for the end-to-end pipeline of
mathematical modeling competitions (such as MCM/ICM and CUMCM):
reading the problem, formulating models, implementing and running
solvers, generating figures, and producing a competition-style report.
The framework comprises \textcolor{blue}{4} specialized ReAct agents---a
\emph{Coordinator}, a \emph{Modeler}, a \emph{Coder}, and a
\emph{Writer}---and a shared on-disk workspace. The Coordinator
occupies the center: it asynchronously dispatches tasks to the \textcolor{blue}{3}
workers, which cooperate \emph{indirectly} through the workspace
rather than by conversing with one another.\footnote{We use
\emph{artifact} to denote any persistent output that one stage
produces and another consumes.}
Figure~\ref{fig:system_overview} shows the architecture.

Given a competition prompt, the Coordinator first writes
\texttt{Problem.md} and decomposes the problem into subproblems
$Q_1, \dots, Q_n$. For each subproblem, the Modeler is dispatched to
produce a modeling plan; the Coder is then dispatched to implement,
run, and visualize, producing data files, figures, and metadata
sidecars. The Writer is dispatched incrementally and adds report
sections as results become available; \S\ref{sec:writer} gives the
order. \textcolor{blue}{By coupling asynchronous role scheduling with
artifact-grounded handoffs, the system allows modeling, computation, and
writing to advance concurrently.}

\subsection{Asynchronous Multi-Type Scheduling}\label{sec:integration}\label{sec:scheduling}

The Coordinator schedules specialized workers rather than executing
stages itself. Each role allows at most one running task and reuses
its subagent instance across dispatches, so that role state and
conversation history persist and a role's context is not interleaved
with a second task of the same role. Figure-generation tasks are the
exception: their calls are stateless, so they run in unbounded
parallelism. \textcolor{blue}{2} primitives implement the policy.
\texttt{dispatch\_task} enqueues a task for a given role and returns
its identifier immediately; \texttt{wait} suspends the Coordinator's
loop until any in-flight task completes and returns that task's
summary.

\begin{figure}[!htbp]
\centering
\includegraphics[width=0.90\linewidth,trim=18bp 34bp 31bp 50bp,clip]{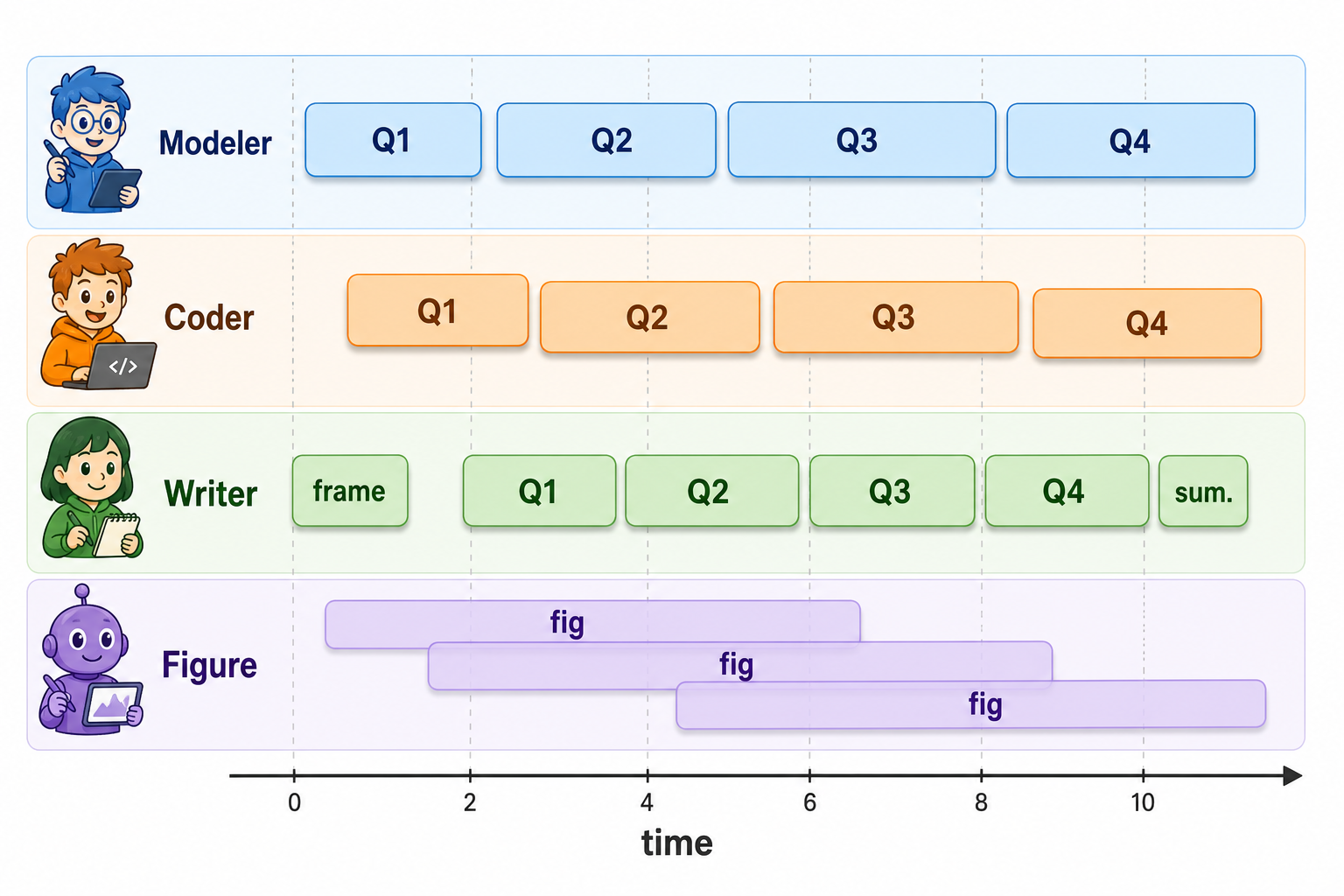}
\caption{Asynchronous scheduling across roles and subproblems. The
Modeler, Coder, and Writer rows allow one running task per role, with
subagent instances reused across dispatches; the Figure row is
unbounded, with multiple \texttt{dispatch\_figure} tasks concurrent.
Time flows left to right, showing how tasks of different roles
overlap across subproblems.}
\label{fig:async_timeline}
\end{figure}

A task is dispatched once its upstream artifacts are
available. For example, the Writer can draft the $Q_1$ section after
the Coder produces $Q_1$'s data and metadata, while the Modeler works
ahead on $Q_3$, \textcolor{blue}{as illustrated in Figure~\ref{fig:async_timeline}}. Different subproblems
can therefore progress through different stages concurrently, and the
report grows alongside the computation rather than after it.
Section~\ref{sec:writer} details the writing passes.

All workers share a uniform tool interface for method
retrieval, sandboxed execution, figure generation, and \LaTeX{}
compilation. Each capability is exposed as a ReAct tool, whether
implemented by local code, an external library such as
HMML~\citep{liu2025mmagent} (\texttt{search\_modeling\_method}), or a
delegated pipeline such as PaperBanana~\citep{zhu2026paperbanana}
(\texttt{dispatch\_figure}). Stable tool signatures allow backends to
be replaced without changing agent prompts, keeping the scheduling
layer independent of backend implementations.

\subsection{Artifact-Grounded Communication}\label{sec:communication}

A naive design forwards each agent's full ReAct trace to the next
agent. \textcolor{blue}{The accumulated context can then exceed the available capacity}, because a Modeler comparing \textcolor{blue}{2}
formulations can emit tens of thousands of tokens. Worse, intermediate
reasoning is misread as final decisions: a trace mixes deliberation
with conclusions and is often truncated, so the Writer may report that
``we adopted formulation $A$'' when $A$ and $B$ were still under
comparison and $B$ was ultimately selected. Both failures follow from
sharing raw conversation rather than curated handoffs.

\textcolor{blue}{\sysname{} replaces shared conversation histories with a \textcolor{blue}{3}-layer
communication mechanism. Each layer provides a distinct type of information
needed by the downstream agent.} The first is a structured digest,
which tells the next worker what has already been decided. The
Coordinator distills each completed task into the decisions,
constraints, and rejected alternatives that the worker needs, and
injects only that digest into the worker's prompt; the full trace is
not forwarded. Conversational state therefore remains isolated per
role, and the report inherits decisions without inheriting the debate.

The second handoff is the \texttt{README} index, which tells the
worker where the artifacts are and which to read. Each worker
maintains a \texttt{README.md} in its subworkspace that indexes its
outputs and explains what each file means and what to read next.
Because the producer writes the index, the downstream agent locates
and understands upstream artifacts through that account rather than
through a global directory listing, and the report can be assembled
from the right files.

\begin{figure}[!htbp]
\centering
\includegraphics[width=0.88\linewidth,trim=49bp 14bp 107bp 0bp,clip]{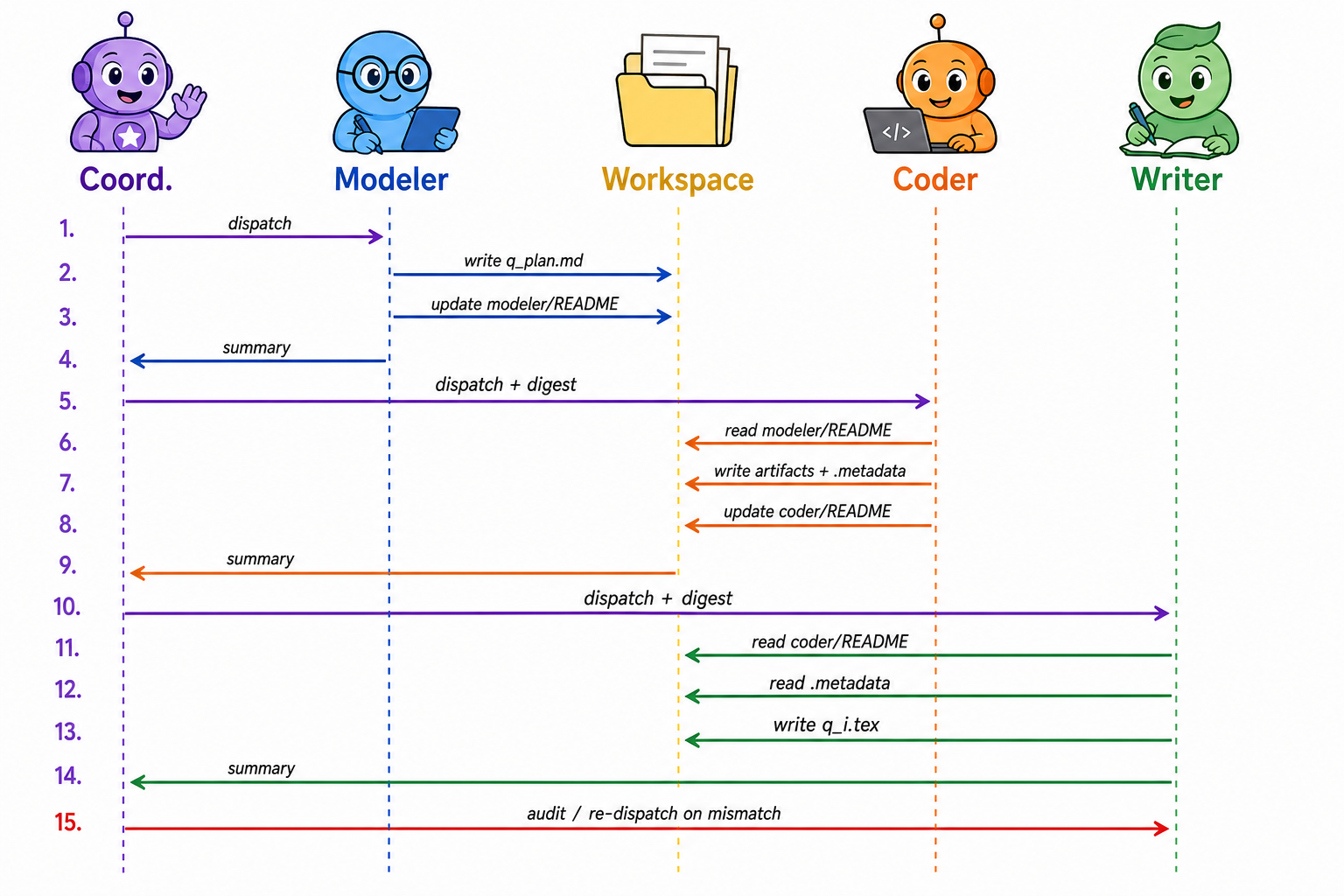}
\caption{End-to-end inter-agent information flow for one subproblem.
The \texttt{Workspace} lane records all file I/O; \texttt{README}
indices and \texttt{.metadata} sidecars let each downstream agent
locate and ground its work.}
\label{fig:contract_seq}
\end{figure}

The third handoff is the metadata sidecar, which tells the Writer
which computed field supports a claim. For each data or figure
artifact, the Coder's execution template generates a
\texttt{.metadata} file (\S\ref{sec:coder}); unlike the artifact
itself, the sidecar is not written by the LLM that produced the
source. It records what the artifact is, which program produced it,
which inputs it consumes, and, for figures, insights read back from
the rendered image. The Writer quotes those fields when reporting a
numerical result, so the sentence and the computation behind it stay
paired.

\begin{figure}[!htbp]
\centering
\begin{minipage}[t]{0.47\linewidth}
\begin{tcolorbox}[title=\texttt{coder/README.md} (excerpt),
  colback=indexbg, colframe=reaslab, fonttitle=\bfseries\sffamily,
  boxrule=0.6pt, arc=1pt, left=4pt, right=4pt, top=3pt, bottom=3pt]
\small\ttfamily
\# Coder outputs for Q2 (SEIR)\\[2pt]
\#\# Files\\
- solve\_q2.py\\
\quad-- SEIR with time-varying R0\\
- data/seir\_curve.csv\\
- figure/seir\_plot.png\\[2pt]
\#\# For Writer\\
- See seir\_curve.csv.metadata\\
\quad for peak\_day / peak\_I (Insights)\\
- See seir\_plot.png.metadata\\
\quad for R0-sensitivity description.
\end{tcolorbox}
\end{minipage}\hfill
\begin{minipage}[t]{0.47\linewidth}
\begin{tcolorbox}[title=\texttt{seir\_curve.csv.metadata},
  colback=indexbg, colframe=reaslab, fonttitle=\bfseries\sffamily,
  boxrule=0.6pt, arc=1pt, left=4pt, right=4pt, top=3pt, bottom=3pt]
\small\ttfamily
\#\#\# Metadata for seir\_curve.csv\\[2pt]
- Description: daily SEIR\\
\quad compartment counts over a\\
\quad 180-day outbreak horizon\\
\quad (piecewise R0 schedule).\\
- Program: solve\_q2.py\\
- Headers: day, S, E, I, R\\
- Predecessors:\\
\quad-- data/population.json\\
\quad-- data/contact\_matrix.csv\\
- Insights:\\
\quad peak\_day: 47\\
\quad peak\_I: 18420\\
\quad final\_R\_pct: 0.487
\end{tcolorbox}
\end{minipage}
\caption{An instance of the \textcolor{blue}{3}-layer contract from a \sysname{} run
on a SEIR subproblem. \emph{Left:} the upstream
\texttt{coder/README.md}, which indexes the Coder's artifacts and
tells the Writer which metadata fields to read. \emph{Right:} the
\texttt{.metadata} sidecar it indexes, generated by code from the
corresponding \texttt{.csv}.}
\label{fig:contract_example}
\end{figure}

\textcolor{blue}{Figure~\ref{fig:contract_seq} illustrates the end-to-end handoff
process for a single subproblem, while Figure~\ref{fig:contract_example}
provides a detailed example of the Coder-to-Writer handoff.} The Coordinator dispatches the
Modeler, which writes a plan and updates \texttt{modeler/README.md};
the Coder reads that plan, implements and runs \texttt{solve\_q2.py},
and emits \texttt{seir\_curve.csv} together with
\texttt{seir\_curve.csv.metadata}. The sidecar records
\texttt{peak\_day=47}, the producing program, and the input files;
\texttt{coder/README.md} tells the Writer to read that field. The
Writer's sentence ``infections peaked on day~47'' therefore has a
traceable chain: report sentence $\rightarrow$ \texttt{peak\_day=47}
$\rightarrow$ \texttt{seir\_curve.csv} $\rightarrow$
\texttt{solve\_q2.py} and its inputs. The Coordinator's audit
(\S\ref{sec:coordinator}) returns any statement that does not match
its sidecar excerpt to the Writer for revision.
Table~\ref{tab:ablation} isolates metadata grounding and independent
dispatch; together with this example, it shows that the \textcolor{blue}{3} handoffs
are complementary rather than redundant.

\subsection{Availability and Deployment}\label{sec:deployment}

\sysname{} is available through \textcolor{blue}{2} complementary paths. The hosted
deployment at ReasLab\footnote{ReasLab can be accessed at
\texttt{https://reaslab.io}.} provides a persistent project workspace
where users can upload problem materials, interact with the agent,
inspect intermediate artifacts, and retrieve compiled reports without
installing the full execution stack. For local use, inspection, and
reuse, we release the ReasLab ReasModel Codex
Bundle,\footnote{The ReasLab ReasModel Codex Bundle can be
accessed at
\url{https://github.com/reaslab/MathModelingAgent-CodexBundle}.}
which packages the modeling workflow for the Codex environment.

We report a production snapshot as evidence of real-world adoption.
From 14 January to 24 August 2026 (UTC), excluding administrator
accounts, 381 users invoked the agent in 763 projects and produced 129
delivered report PDFs across 123 projects.\footnote{A user turn is a
chat record with \texttt{role=USER} and
\texttt{agentId=math-modeling}; the window contains 8,836 turns. The
storage scan deduplicated repeated deliveries of the same report, and
116 of the 129 reports were linked to non-administrator activity.}
These figures come from a single production cluster and reflect
artifacts present at snapshot time. They measure adoption and
retention, not task success, and are reported separately from the
benchmark scores in Section~\ref{sec:experiments}.

\section{Agent Design}\label{sec:agents}

Section~\ref{sec:system} described how the roles coordinate. This
section describes what each contributes to the chain from a modeling
decision to an executable computation, from computation to evidence,
and from evidence to a complete report. The Coordinator keeps the
chain moving; the Modeler, Coder, and Writer own its \textcolor{blue}{3} links. The
\textcolor{blue}{4} workflow figures give the operating order for each role.

\subsection{Coordinator}\label{sec:coordinator}

The \texttt{Coordinator} drives asynchronous progress through \textcolor{blue}{3} main activities: problem
decomposition, structured handoffs, and report auditing; the
concurrency policy itself is described in
Section~\ref{sec:integration}. It holds the dispatch primitives and
decides when the report is ready to leave the loop. It receives
completed work, distills each result into a digest for the next
worker, and audits the Writer's claims against the artifacts.

\begin{figure}[!htbp]
\centering
\includegraphics[width=0.86\linewidth,trim=67bp 29bp 79bp 7bp,clip]{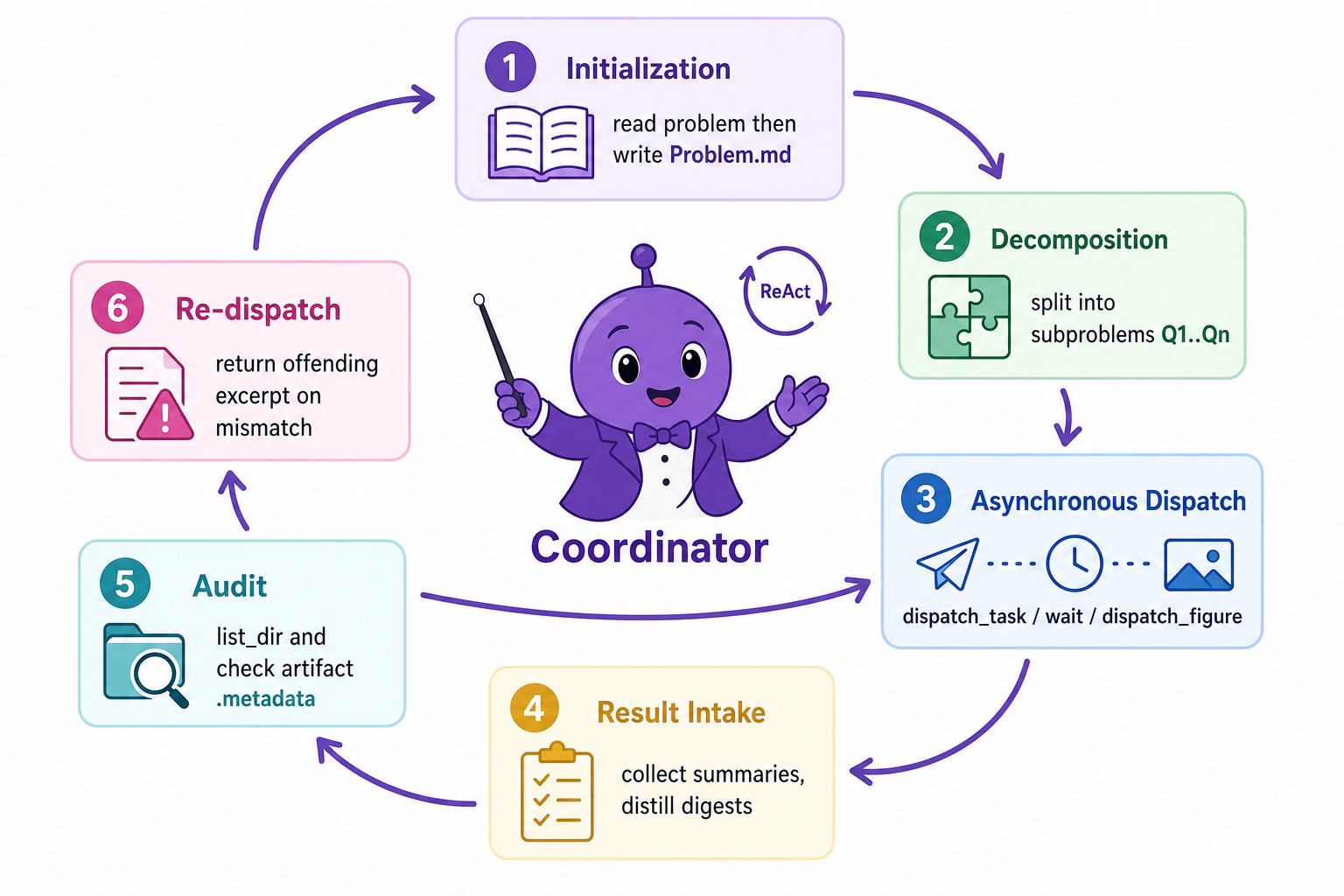}
\caption{Workflow of the Coordinator. The Coordinator is the only
agent in \sysname{} that holds the dispatch primitives; it iterates
through decomposition, dispatch, intake, audit, and re-dispatch
stages until the report is grounded against every artifact's
\texttt{.metadata}.}
\label{fig:coordinator_workflow}
\end{figure}

First, the Coordinator fixes the unit of work through problem decomposition. It reads the competition prompt and any attachments, writes
\texttt{Problem.md}, and splits the problem into subproblems
$Q_1,\dots,Q_n$. Each dispatch is then tied to one of these units, so
a completed task can be attributed to a subproblem and handed to the
next role without renegotiating scope.

Second, the Coordinator uses a digest for each conversational handoff.
When a worker returns, it distills the result
into the decisions, constraints, and rejected alternatives that the
next worker needs, and injects only that digest into the next prompt.
The next worker starts from settled decisions rather than from the
previous worker's full trace; this is what keeps contexts bounded
while preserving the decisions that matter.

Third, report auditing closes the loop. Before the report leaves
the system, the Coordinator checks its quantitative claims against the
corresponding \texttt{.metadata} sidecars; a mismatch is returned to
the Writer with the offending excerpt and the metadata block, and the
Writer revises. A report is finished only when its numbers survive
this check.

\subsection{Modeler}\label{sec:modeler}

The \texttt{Modeler} turns an underspecified subproblem into a
mathematical specification that the Coder can implement: a justified
method choice, decision variables, an objective, constraints, and a
plan.

\begin{figure}[!htbp]
\centering
\includegraphics[width=0.90\linewidth,trim=17bp 83bp 14bp 53bp,clip]{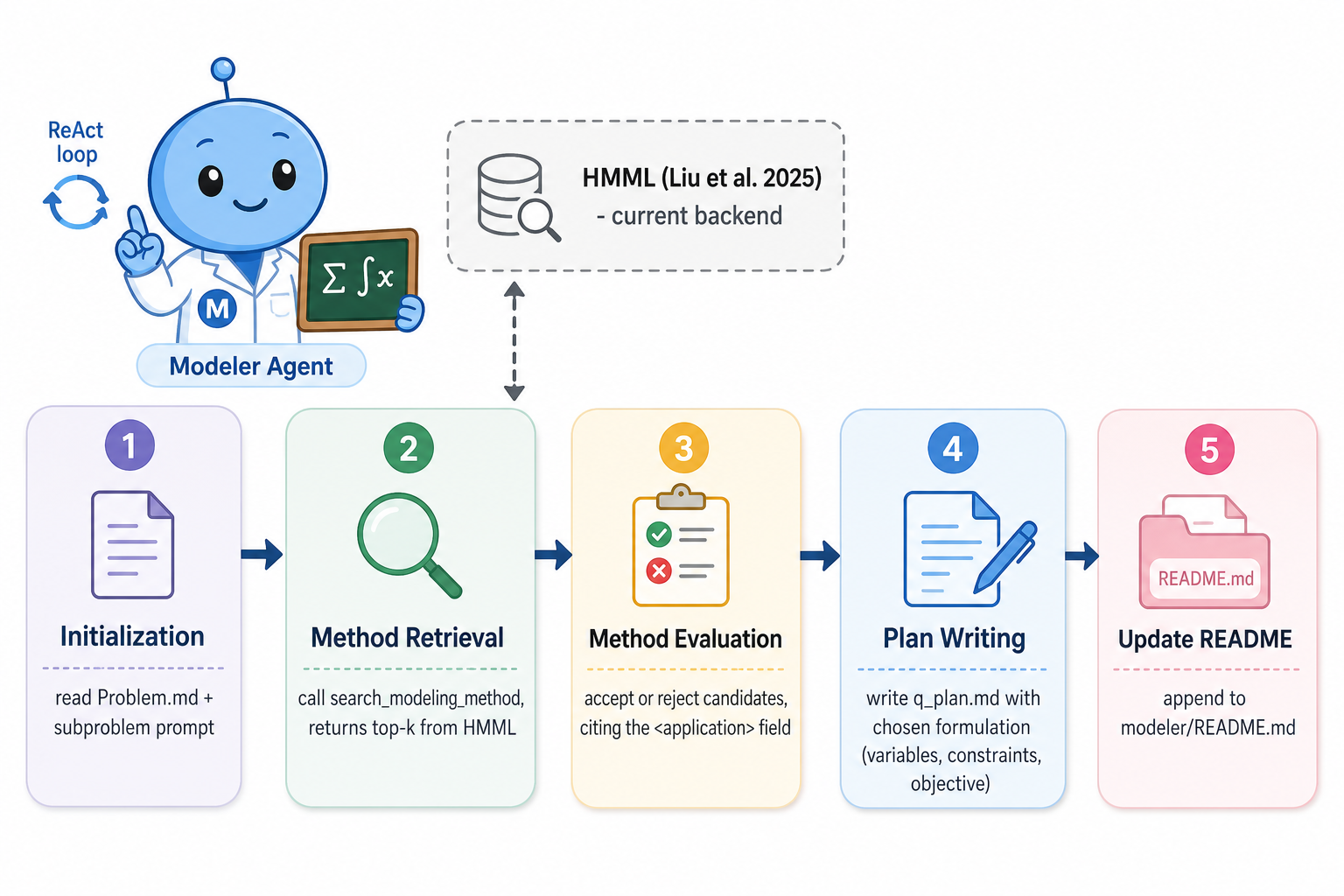}
\caption{Workflow of the Modeler. The Modeler invokes the HMML library
exposed as a single ReAct tool, evaluates retrieved candidates with
citation-grounded justifications, and emits a formal modeling plan
into the workspace.}
\label{fig:modeler_workflow}
\end{figure}

To select candidate methods, the Modeler queries \texttt{search\_modeling\_method} with a query
derived from the subproblem. The query is built from the subproblem
statement and the Coordinator's digest, so retrieval is conditioned
on decisions already made. The tool returns the top-$k$ HMML nodes
with their hierarchy paths and \texttt{<modeling\_method>} /
\texttt{<core\_idea>} / \texttt{<application>} schema. The Modeler
does not simply take the top hit: it accepts or rejects each
candidate. Each rejection quotes the candidate's
\texttt{<application>} field---the condition under which the method
applies---and explains its mismatch with the current subproblem.
Method selection is therefore tied to the method's stated scope, and
the resulting decision trail is auditable.

Once a method is chosen, the Modeler writes \texttt{q\_plan.md}. The
plan records the chosen method and its hierarchy path, a quoted
fragment of its schema, the formal model (decision variables,
objective, constraints), and a ledger of rejected candidates with
their citations. The plan is more than a method name: the variables
and constraints are the interface between modeling and code, and the
rejection ledger records why alternatives were not used. This is the
object the Coder implements; it is also the description of
assumptions and model structure that the Writer later reports.

HMML itself originates from LLM-MM-Agent~\citep{liu2025mmagent};
\sysname{}'s contribution is the calling contract around it---a fixed
tool signature, at least one retrieval per subproblem, and a citation
obligation on rejection---rather than the method catalog. The chosen
method family also determines which solver class the Coder will use.
The Modeler's output is a specification the Coder can implement
without re-deriving the model.

\subsection{Coder}\label{sec:coder}

The \texttt{Coder} turns the plan into a running program and a
traceable record of what that program produced. It is where a
mathematical specification becomes a computation that the report can
cite.

\begin{figure}[!htbp]
\centering
\includegraphics[width=0.90\linewidth,trim=24bp 34bp 30bp 32bp,clip]{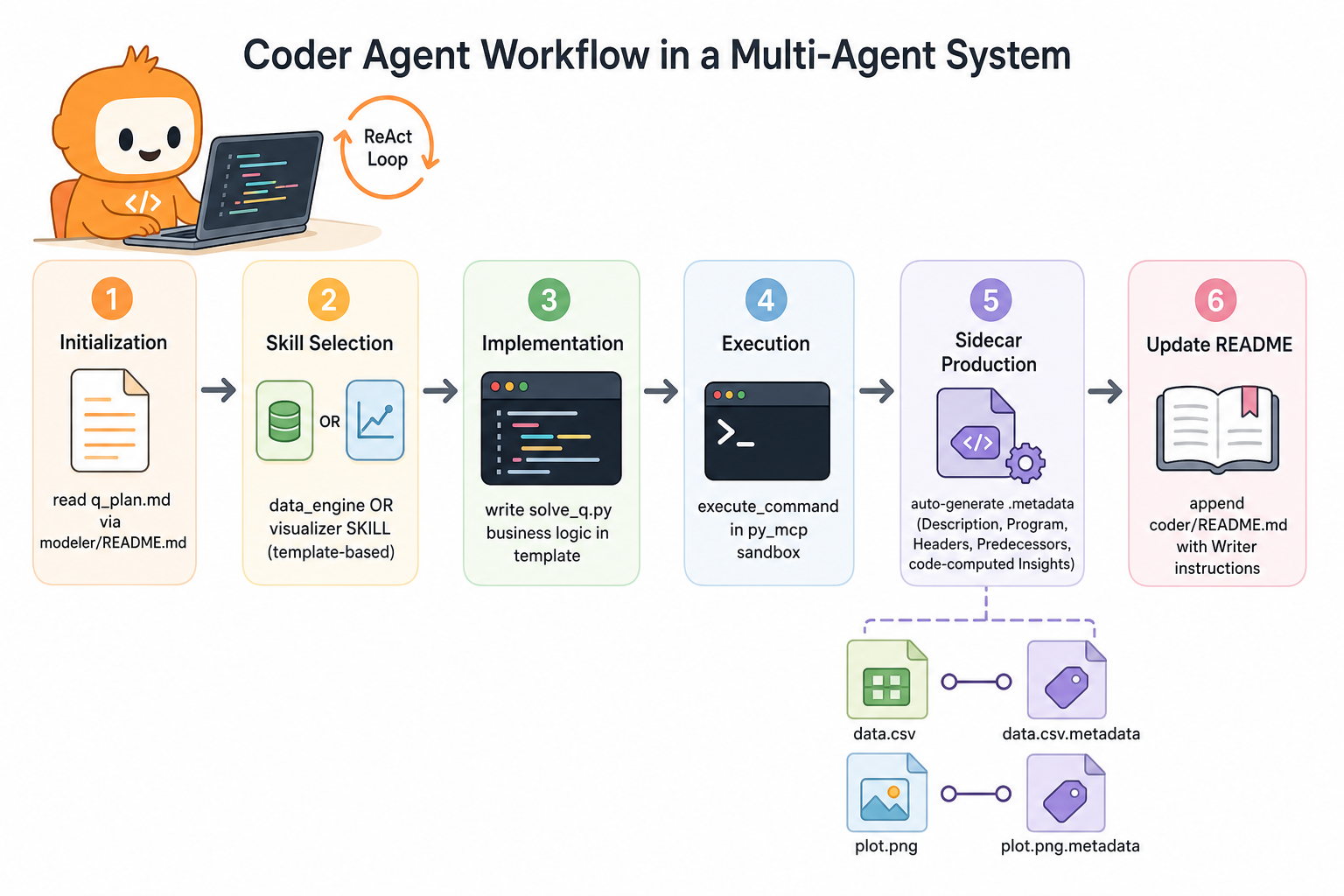}
\caption{Workflow of the Coder. Each artifact (\texttt{.csv},
\texttt{.png}, \texttt{.txt}, \texttt{.json}) is produced together
with a paired \texttt{.metadata} sidecar---written by code, not by
the LLM---to enforce the anti-hallucination contract of
\S\ref{sec:communication}.}
\label{fig:coder_workflow}
\end{figure}

To couple evidence generation with execution, each Coder task starts from a skill-specific \texttt{template.py} with \textcolor{blue}{3}
blocks: a configuration block, a business-logic block, and an
immutable main block. The Coder writes only the business logic. The
skill prompt declares the main block immutable, and the block itself
appends a \texttt{<artifact>.metadata} sidecar after the program
runs. Metadata generation is thus embedded in the execution template
and runs alongside artifact production.

The Coder selects the \texttt{data\_engine} or \texttt{visualizer}
skill according to the task, writes the program, and executes it in
the \texttt{py\_mcp} sandbox. A data artifact leaves with a sidecar
recording a description, the producing program, column headers, and
predecessor inputs. For a table, the sidecar records the column
headers in the order in which they appear; for a figure, a separate
vision-LLM pass re-reads the rendered PNG and writes an
\emph{Insights} block reporting the few statistics that support what
the picture shows. The resulting insights are generated from the
actual rendered image. The Coder then appends a per-artifact entry to
\texttt{coder/README.md}, naming the fields the Writer should quote.

For specialist solving, \sysname{} does not add a dedicated ReAct
tool. The sandbox image provides bindings for Gurobi, cvxpy,
\texttt{scipy.optimize}, and OR-Tools, and the \texttt{data\_engine}
skill declares the solver inventory together with an invocation
pattern for each family. The Modeler's method choice thus maps to a
concrete solver without capability discovery inside the Coder's loop.
Because the inventory lives in the skill, the method-to-solver
mapping can be updated without changing the agent-facing tool list.
Licenses and environment configuration stay in the sandbox and never
enter the LLM context.

The Coder's output is therefore paired: an artifact and a sidecar
that says what it is, which program produced it, and which inputs it
depends on. The Writer can cite the sidecar; the Coordinator can
check it. This pairing is what lets the Writer produce a quantitative
sentence without reading the code, and lets the Coordinator audit it
without rerunning the experiment. It is the producer side of the
grounding contract in \S\ref{sec:communication}.

\subsection{Writer}\label{sec:writer}

The \texttt{Writer} turns artifacts into a report whose claims remain
tied to the computation. It is the system's primary metadata consumer,
and the role where the report-native objective is realized.

\begin{figure}[!htbp]
\centering
\includegraphics[width=0.94\linewidth,trim=16bp 209bp 21bp 117bp,clip]{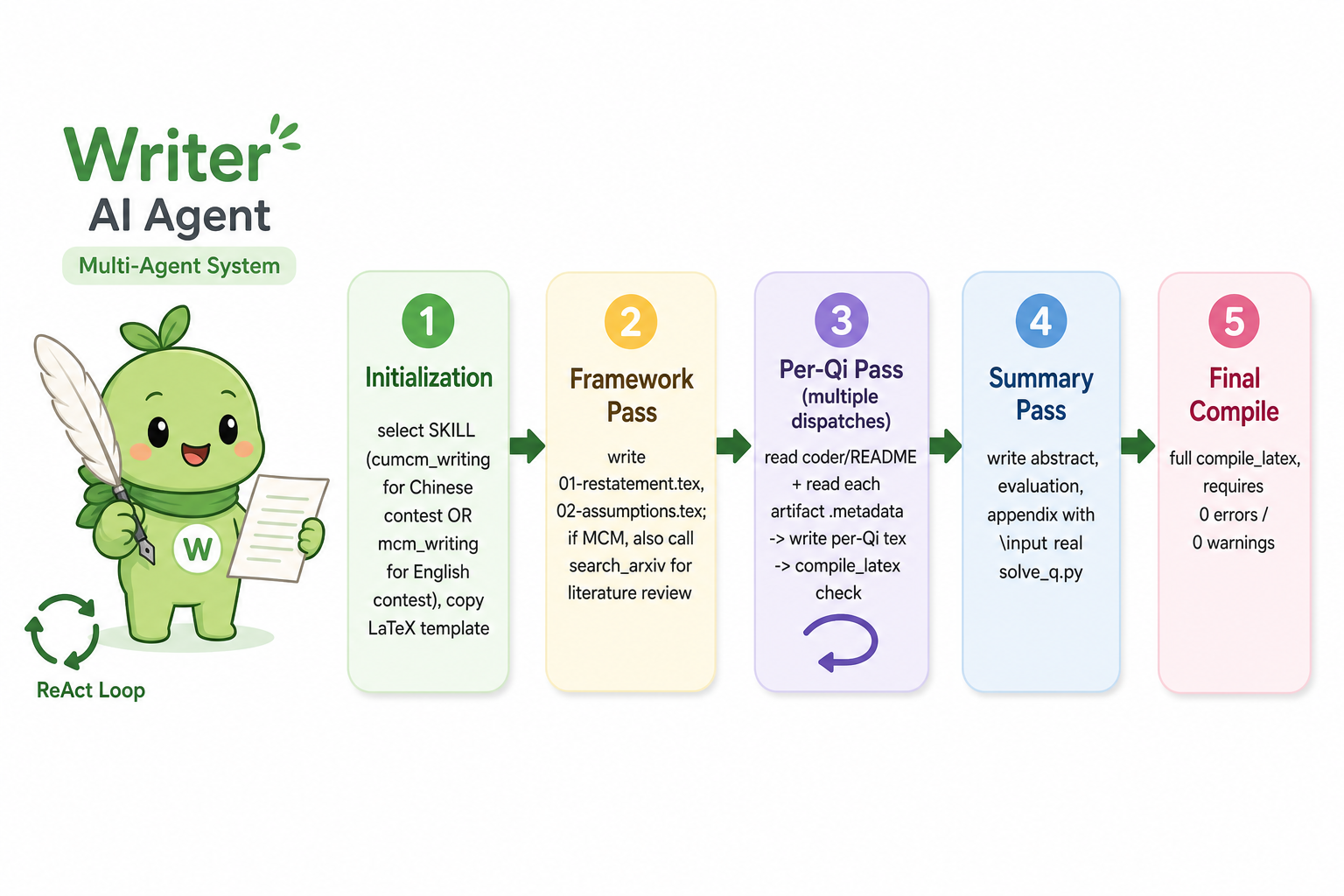}
\caption{Workflow of the Writer. Framework chapters, per-subproblem
sections, and closing chapters are dispatched in that order, each
followed by a \texttt{compile\_latex} gate.}
\label{fig:writer_workflow}
\end{figure}

Writing proceeds in passes to integrate evidence progressively. The Writer first drafts the framework chapters---problem
restatement and assumptions, plus a literature review for English
contests---which do not depend on results. As each Coder task
completes, it adds the corresponding per-subproblem section, reading
\texttt{coder/README.md} to locate the artifacts and the indexed
\texttt{.metadata} files to ground each statement. It closes with the
abstract, the evaluation section, and the appendix. This ordering
lets writing start before results exist and finish as the last result
arrives.

To enforce metadata grounding, the Writer's prompt forbids describing an artifact whose \texttt{.metadata} has
not been read in the current ReAct trace, and quantitative claims
must quote the corresponding fields. This is how the grounding
contract becomes prose: a sentence about a peak or a bound is written
from a field, not from memory. When a required field is missing, the
Writer must either locate another artifact or leave the claim out.

\textcolor{blue}{2} further choices protect the final document. The appendix
\texttt{\textbackslash input}-includes the actual
\texttt{solve\_q*.py} files, so the program shown in the report
corresponds to the program that ran. \texttt{compile\_latex} runs
after each dispatch; a per-section failure is caught before later
sections depend on it, and the final build must have zero errors and
zero warnings. Contest-specific skills (\texttt{cumcm\_writing} for
Chinese CUMCM, \texttt{mcm\_writing} for English MCM/ICM) set the
chapter order and voice. \textcolor{blue}{Once the report is complete, the Writer
returns it to the Coordinator for audit.}
The report is finished only when its quantitative statements survive
that check and still point to the artifacts behind them.

\section{MathModeling-Bench}
\label{sec:cumcm-bench}

Standard mathematical-reasoning benchmarks primarily evaluate short-form answers or derivations~\citep{he2024olympiadbench,lu2024mathvista}. In contrast, mathematical modeling requires a system to interpret an underspecified real-world problem, process heterogeneous attachments, formulate and solve a model, validate its conclusions, and communicate the resulting evidence in a coherent report. These outputs are long-form and non-unique, which makes exact-match evaluation inappropriate. We therefore construct \textsc{MathModeling-Bench}, a benchmark of 46 authentic contest problems for evaluating end-to-end mathematical modeling. The benchmark preserves original competition artifacts to retain authentic task complexity and uses problem-specific rubrics to evaluate open-ended modeling reports.

\subsection{Task Formulation}
\label{sec:cumcm-task}

We represent benchmark instance $i$ as
\begin{equation}
    \mathcal{B}_i = (P_i, A_i, R_i),
\end{equation}
where $P_i$ is the original problem statement, $A_i$ denotes \textcolor{blue}{the data and supporting files distributed with that contest problem (empty if none are supplied)}, and $R_i$ is an evaluator-only, problem-specific rubric. The evaluated system observes $(P_i,A_i)$ and produces a complete modeling report $Y_i$; it never observes $R_i$. A successful report is expected to state assumptions, formulate the mathematical model, describe the solution procedure, report and interpret results, and provide validation or sensitivity analysis when required by the task. This separation preserves the natural task interface while allowing the evaluator to distinguish a genuinely task-complete solution from a plausible but generic modeling narrative.

\subsection{Benchmark Construction}
\label{sec:cumcm-construction}

The current benchmark contains 46 problems from the China Undergraduate Mathematical Contest in Modeling (CUMCM), spanning 2011--2025.\footnote{CUMCM official website: \texttt{https://www.mcm.edu.cn}.} We retain the original problem statements rather than converting them into a text-only surrogate: 27 are distributed as DOCX files and 19 as PDF files. \textcolor{blue}{34} problems include standalone attachments, yielding 96 top-level attachment items in total. These artifacts include spreadsheets, CSV files, documents, raster images, and video, and thus require systems to reason over both narrative specifications and external data.

Each benchmark task contains the original statement and an attachment directory; the corresponding rubric is stored separately for evaluation. We validate the active paths for all statements, attachment directories, and rubric files before evaluation. Review materials and derived rubrics are excluded from the solver-visible package. This access boundary is essential because the rubric may disclose expected analyses, constraints, or numerical checks that are not stated explicitly in the problem.

\subsection{Problem-Specific Rubrics}
\label{sec:cumcm-rubric}

Open-ended modeling problems admit multiple defensible formulations, yet each problem also contains task-specific obligations that a report must address. A single generic report-level rubric cannot reliably detect omissions such as ignoring a supplied dataset, dropping a physical constraint, failing to provide a requested decision, or answering only a simplified surrogate problem. We therefore associate each problem with a structured set of requirements
\begin{equation}
    R_i = \{r_{ij}\}_{j=1}^{n_i}, \qquad
    r_{ij}=(\mathrm{id}_{ij},g_{ij},t_{ij},w_{ij}),
\end{equation}
where $g_{ij}$ identifies a subproblem or criterion group, $t_{ij}$ is a self-contained requirement, and $w_{ij}>0$ is its percentage weight. For every problem, $\sum_j w_{ij}=100$.

For the current benchmark collection, rubric construction begins from publicly available official problem-specific evaluation criteria.\footnote{CUMCM official website: \texttt{https://www.mcm.edu.cn}.} A source-grounded language model first produces a structured draft containing source quotations, line spans, criterion groups, and raw weights. The draft is then manually reviewed and revised to correct extraction errors, merge overly fragmented criteria, preserve valid alternative solution paths, and ensure that each requirement can be judged independently. Finally, a deterministic formalizer removes drafting metadata, normalizes the reviewed weights to 100, and validates the result against a fixed JSON schema. The resulting 46 rubrics contain 282 requirements. Importantly, alternative methods are encoded conditionally: a report is not penalized for omitting an unchosen alternative when it implements another accepted path correctly. The formal schema is independent of rubric provenance; problems without official review material can consequently be incorporated through generated drafts followed by expert review under the same representation.

\subsection{Multi-Dimensional Evaluation}
\label{sec:cumcm-evaluation}

We evaluate each report along \textcolor{blue}{3} complementary dimensions: Problem-Specific Requirement Fulfillment (PSRF), Modeling Solution Quality (MSQ), and Report Visual Quality (RVQ). The first measures problem-conditioned fulfillment, while the latter \textcolor{blue}{2} capture cross-problem scientific and communication quality.

To compute PSRF, for each applicable requirement $r_{ij}$, an evidence-grounded judge
assigns a normalized fulfillment score $s_{ij}\in[0,1]$ and records
supporting evidence from the report. Let $\mathcal{J}_i$ denote the set
of requirements applicable to report $Y_i$. The corresponding weights
are normalized over the applicable requirements such that
$\sum_{j\in\mathcal{J}_i} w_{ij}=1$. The PSRF score is

\begin{equation}
S_i^{\mathrm{PSRF}}
=
100\sum_{j\in\mathcal{J}_i} w_{ij}s_{ij}.
\label{eq:cumcm-psrf}
\end{equation}
The judge receives the original problem statement, the hidden rubric, and the generated report. Full credit requires substantive support from the model, results, and reasoning; a vague mention does not suffice. Missing judgments are assigned zero deterministically. For data- or computation-dependent claims, an exact-looking value or a statement that code was executed is not itself accepted as evidence: the report must provide a traceable connection from inputs and assumptions to outputs. Unsupported upstream results also reduce credit for dependent forecasts, decisions, or sensitivity analyses.

Whereas PSRF measures problem-specific fulfillment, MSQ evaluates task-agnostic modeling rigor using \textcolor{blue}{6} dimensions: problem understanding, solution correctness, validation and sensitivity, novelty and contribution, structure and logic, and academic writing. Their weights are $(0.20,0.25,0.20,0.10,0.15,0.10)$, respectively. If $m_{ik}\in[0,100]$ is the score for dimension $k$, then
\begin{equation}
    S_i^{\mathrm{MSQ}} = \sum_{k=1}^{6}\alpha_k m_{ik}.
    \label{eq:msq}
\end{equation}
The MSQ judge observes the report and a compact evidence summary from PSRF. It recomputes the weighted total from the \textcolor{blue}{6} dimension scores, rather than accepting an unconstrained holistic score. Purely visual defects are excluded here unless they obstruct mathematical interpretation.

Finally, RVQ measures whether the rendered report communicates the modeling argument effectively. The evaluator considers figure and table relevance, labels and units, formula readability, captions and cross-references, result presentation, and document organization; decorative graphics receive no credit. \textcolor{blue}{For PDF reports, the evaluator renders up to 15 representative pages, prioritizing pages with figures and tables while also including some pages with formulas, and performs page-grounded multimodal review.} If page rendering is unavailable, it falls back to source text and document metadata and records the lower-confidence review mode. The resulting score is denoted $S_i^{\mathrm{RVQ}}\in[0,100]$.

We use a weighted additive aggregation to reflect the priorities of
end-to-end mathematical modeling. PSRF receives the largest weight
($0.50$) because fulfilling the problem-specific requirements is the
primary criterion: a report that omits a required analysis, constraint,
or decision should not receive a high overall score solely because it is
well written. MSQ receives a weight of $0.35$ to capture modeling
correctness, rigor, validation, and scientific communication across
different problems. RVQ receives a weight of $0.15$ because clear visual
presentation is important for communicating results, but should not
compensate for missing or incorrect substantive work. The \textcolor{blue}{3} component
scores are reported together with the aggregate so that systems with
different strengths are not reduced to a single number.
\begin{equation}
    S_i = 0.50S_i^{\mathrm{PSRF}}
        + 0.35S_i^{\mathrm{MSQ}}
        + 0.15S_i^{\mathrm{RVQ}}.
    \label{eq:cumcm-overall}
\end{equation}
These weights are fixed protocol choices that encode the benchmark's
evaluation priorities, rather than claims of a uniquely optimal utility
function. 

\subsection{Benchmark Statistics}
\label{sec:cumcm-statistics}

\begin{figure}[t]
    \centering
    \includegraphics[width=\linewidth]{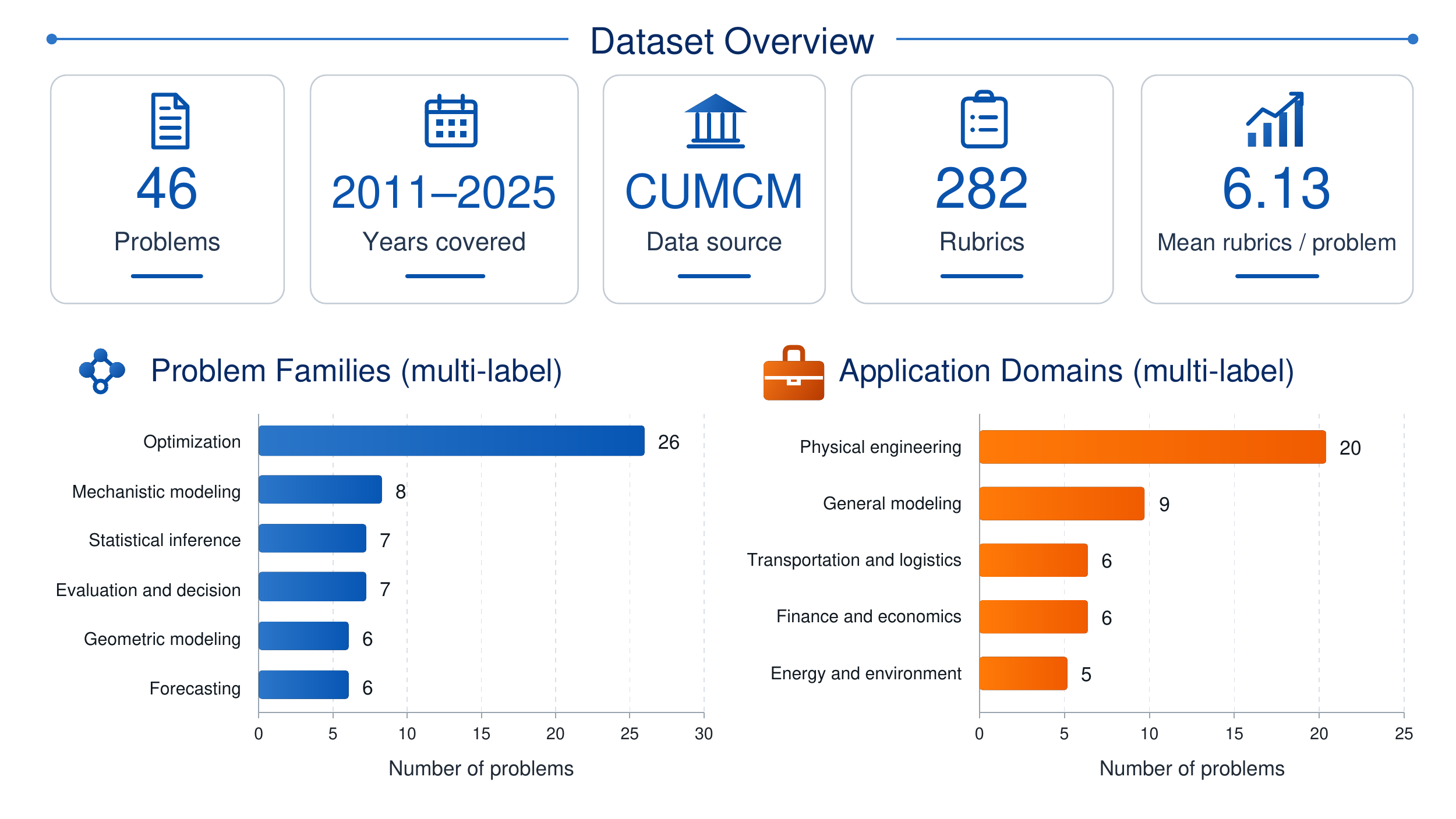}
    \caption{Dataset overview of 46 problems from the China Undergraduate Mathematical Contest in Modeling (CUMCM), spanning 2011--2025. The figure summarizes the data source, rubric statistics, and the distributions of problem families and application domains. The dataset contains 282 rubrics, with a mean of 6.13 rubrics per problem. Categories are multi-label; counts do not sum to 46.}
    \label{fig:cumcm-statistics}
\end{figure}

Figure~\ref{fig:cumcm-statistics} summarizes the collection. \textcolor{blue}{Each problem contains between 2 and 12 rubric requirements.} The task taxonomy is multi-label: optimization is prominent, but the benchmark also includes mechanistic modeling, statistical inference, evaluation and decision, geometric modeling, and forecasting. Application domains similarly span physical engineering, transportation and logistics, finance and economics, and energy and environment. These distributions provide diverse rather than balanced coverage; category counts therefore should not be interpreted as a stratified sampling design.

\section{Experiments}
\label{sec:experiments}

We evaluate \sysname{} on MathModeling-Bench and MM-Bench, and ablate
metadata grounding and independent worker dispatch. \textcolor{blue}{We compare against
vanilla GPT-5.6-Sol~\citep{openai2026gpt56} and DeepSeek
V4.1~Flash~\citep{deepseek2026v4report}, as well as the sequential modeling agent
LLM-MM-Agent~\citep{liu2025mmagent} using the same backbones.}



\subsection{Performance Evaluation on Our Benchmark}
\label{subsubsec:exp-our-benchmark}

We begin with MathModeling-Bench, our benchmark designed specifically for native end-to-end mathematical modeling reports. This benchmark matches the target setting of \sysname: the system receives the original competition statement and attachments, and is evaluated on the quality of the final paper rather than on a converted intermediate representation. Using the PSRF, MSQ, and RVQ framework introduced in Section~\ref{sec:cumcm-bench}, we compare multiple frontier models and agent systems, with the aggregate results reported in Table~\ref{tab:overall_benchmark_results}.

\begin{table}[htbp]
\centering
\caption{Overall Benchmark Results}
\label{tab:overall_benchmark_results}
\begin{tabular}{ll|cccc}
\toprule
\textbf{Method} & \textbf{Backbone} & \textbf{PSRF} &
\textbf{MSQ} &
\textbf{RVQ} &
\textbf{Final Score} \\
\midrule

\multirow{2}{*}{Vanilla LLM}
& DeepSeek V4.1 Flash & 49.77 & 51.68 & 35.46 & 48.29 \\
& GPT-5.6-Sol         & 55.48 & 53.17 & 72.87 & 57.26 \\
\midrule

\multirow{2}{*}{LLM-MM-Agent}
& DeepSeek V4.1 Flash & 45.89 & 47.17 & 53.22 & 47.44 \\
& GPT-5.6-Sol         & 56.11 & 54.84 & 54.90 & 55.48 \\
\midrule

\multirow{2}{*}{\textbf{ReasModel}}
& DeepSeek V4.1 Flash & 72.57 & 64.33 & 80.46 & 70.87 \\
& \textbf{GPT-5.6-Sol}         & \textbf{75.29} & \textbf{73.67} & \textbf{80.56} & \textbf{75.51} \\

\bottomrule
\end{tabular}
\end{table}

Table~\ref{tab:overall_benchmark_results} shows that \sysname{} achieves the strongest overall performance in the report-native setting, reaching a final score of 75.51. Compared with LLM-MM-Agent, \sysname{} attains the best scores across all dimensions. 

\begin{figure}[!htbp]
\centering
\includegraphics[width=0.82\linewidth]{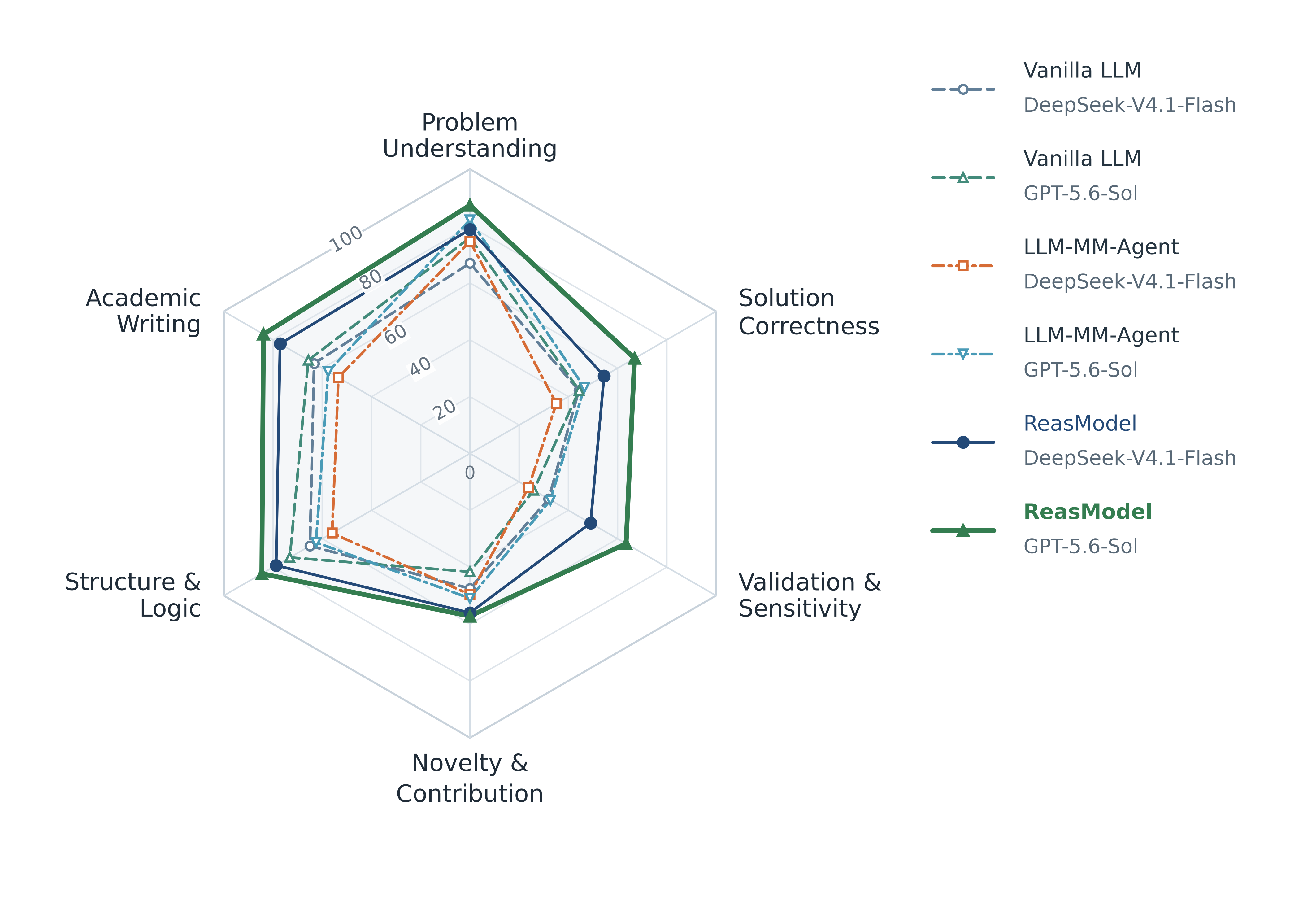}
\caption{Radar-chart comparison of modeling quality across 6 systems and 6 evaluation dimensions. Scores are reported on a 0--100 scale.}
\label{fig:modeling-quality-radar}
\end{figure}

Figure~\ref{fig:modeling-quality-radar} presents the \textcolor{blue}{6} subdimensions of Modeling Solution Quality (MSQ), as defined in Section~\ref{sec:cumcm-bench}: problem understanding, solution correctness, validation and sensitivity, novelty and contribution, structure and logic, and academic writing. Their weights in the aggregate MSQ score are $(0.20,0.25,0.20,0.10,0.15,0.10)$, respectively; the radar chart displays the mean subdimension scores on a 0--100 scale before weighting. ReasModel with GPT-5.6-Sol achieves the highest reported mean on all \textcolor{blue}{6} subdimensions.
\subsection{Performance Evaluation on MM-Bench}
\label{subsubsec:exp-mmbench}

We next evaluate \sysname{} on MM-Bench, an MCM/ICM-derived benchmark for end-to-end mathematical modeling that scores a structured \texttt{solution.json} solution rather than the final report. We compare with a Vanilla LLM and LLM-MM-Agent, and use an adapter prompt to convert \sysname{}'s native outputs into the required JSON format.

\begin{table}[htbp]
\centering
\caption{MM-Bench Evaluation Results}
\label{tab:mmbench_evaluation_results}
\begin{tabular}{ll|ccccc}
\toprule
\textbf{Method} & \textbf{Backbone} &
\textbf{AE} &
\textbf{MR} &
\textbf{PS} &
\textbf{RBA} &
\textbf{Overall} \\
\midrule

\multirow{2}{*}{Vanilla LLM}
& DeepSeek V4.1 Flash & 7.26 & 6.10 & 6.21 & 5.37 & 6.24 \\
& GPT-5.6-Sol & 8.28 & 7.51 & 7.31 & 7.42 & 7.63 \\
\midrule

\multirow{2}{*}{LLM-MM-Agent}
& DeepSeek V4.1 Flash & 7.94 & 6.76 & 6.02 & 7.78 & 7.12 \\
& GPT-5.6-Sol &8.50 &7.91 &6.56 &7.75 &7.68 \\
\midrule

\multirow{2}{*}{\textbf{ReasModel}}
& DeepSeek V4.1 Flash & 8.62 & 7.41 & 7.06 & 8.30 & 7.85 \\
& GPT-5.6-Sol & \textbf{8.95} & \textbf{8.00} &\textbf{7.53} &\textbf{8.31} &\textbf{8.19} \\

\bottomrule
\end{tabular}
\vspace{2pt}

\parbox{0.92\linewidth}{\footnotesize\raggedright
AE: Problem Analysis and Understanding; MR: Modeling Rigorousness; PS: Practicality and Scientificity; RBA: Result and Bias Analysis.
}
\end{table}

\begin{figure}[!t]
\centering
\includegraphics[width=\linewidth]{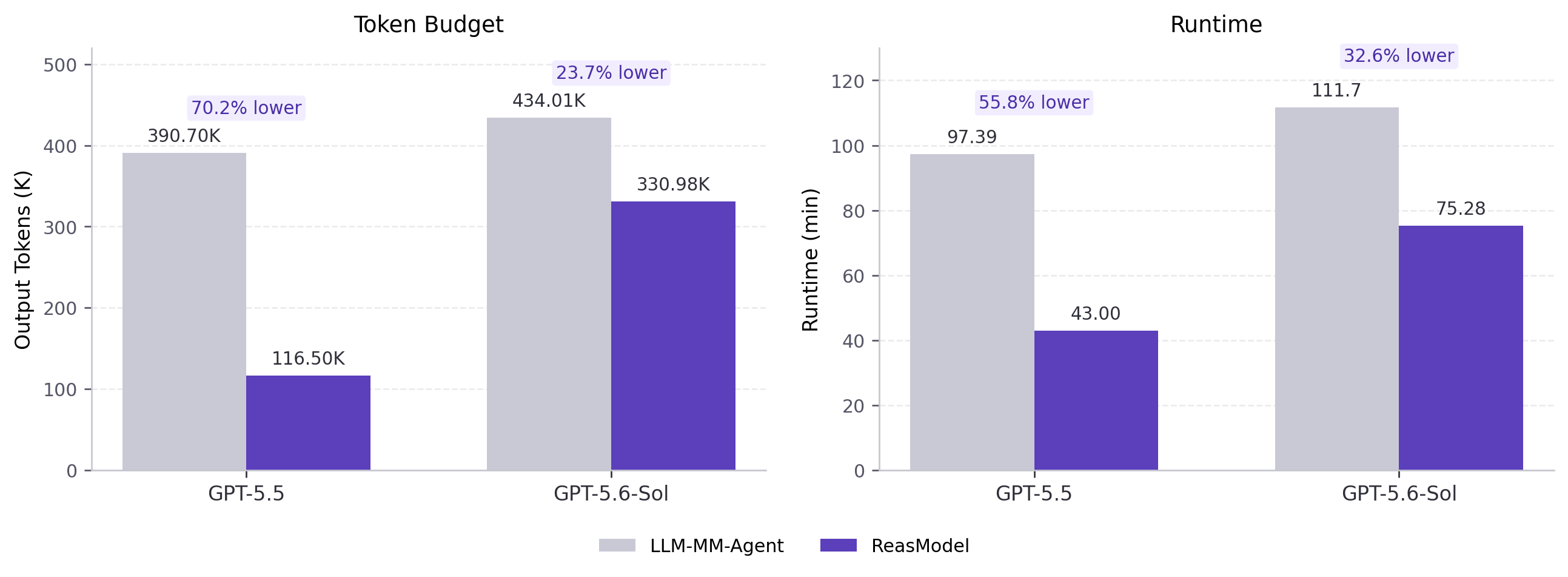}
\caption{Generation cost of complete reports on the 32 MM-Bench problems: output tokens (left) and end-to-end runtime (right), under matched backbones. Under GPT-5.6-Sol, ReasModel uses 23.7\% fewer output tokens and 32.6\% less runtime than LLM-MM-Agent.}
\label{fig:mmbench_resource_budgets}
\end{figure}

\begin{figure}[!t]
\centering
\includegraphics[width=\linewidth]{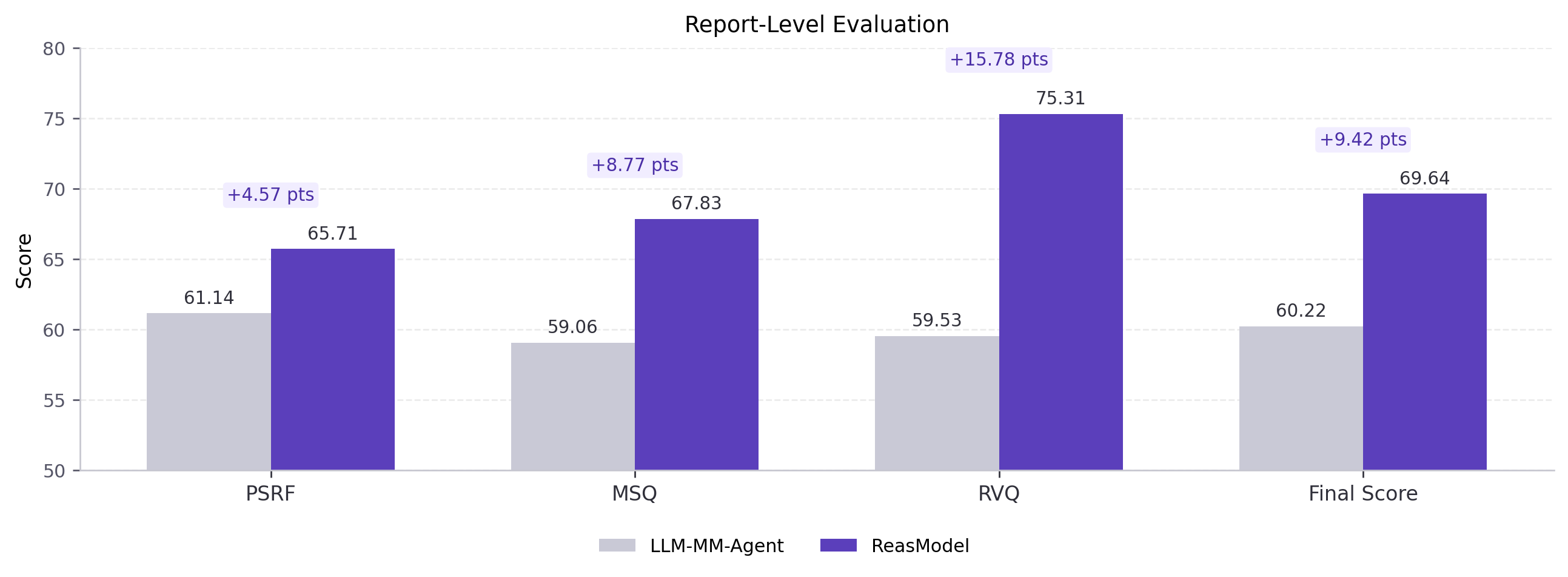}
\caption{Evaluation of the complete reports on the same 32 problems, rescored with the PSRF, MSQ, RVQ, and Final Score of MathModeling-Bench (0--100 scale). Under GPT-5.6-Sol, ReasModel outperforms LLM-MM-Agent on all \textcolor{blue}{4} metrics, with the largest gain in RVQ.}
\label{fig:mmbench_report_evaluation_results}
\end{figure}
\FloatBarrier

As shown in Table~\ref{tab:mmbench_evaluation_results}, \sysname{} attains the highest overall score (8.19) and leads on every dimension. However, MM-Bench scores a converted JSON representation rather than the final report, which limits its view of overall report quality; the conversion may also discard information available in the native outputs.

We therefore let both systems generate complete reports for the same 32 problems and examine them separately from the MM-Bench scores. Figure~\ref{fig:mmbench_resource_budgets} reports the cost of generating these reports under matched backbones; \sysname{} uses fewer output tokens and less runtime than LLM-MM-Agent. We then rescore the complete reports with the PSRF, MSQ, and RVQ criteria of MathModeling-Bench, using the same 0--100 scale as in Section~\ref{subsubsec:exp-our-benchmark}.  As shown in Figure~\ref{fig:mmbench_report_evaluation_results}, \sysname{} with GPT-5.6-Sol outperforms LLM-MM-Agent on all \textcolor{blue}{3} metrics and the final score, with the largest gain in RVQ.

\subsection{Ablation Experiments}
\label{subsec:ablation}

The ablation study uses the same 46 benchmark instances as
Section~\ref{subsubsec:exp-our-benchmark}. All configurations use DeepSeek
V4.1 Flash. Starting from the full system, we remove metadata grounding,
independent worker dispatch, or both, leaving the remaining workflow
unchanged.

In the configuration without metadata, we disable the production and consumption of the \texttt{.metadata} sidecars
specified in Appendix~\ref{app:metadata_schema} but retain role dispatch,
subworkspaces, and \texttt{README.md} indices. The Coder therefore produces
primary artifacts only, and the Writer's numerical and visual descriptions are
no longer gated by sidecar reads; they must instead be derived from the
remaining plans, code, results, and figures. This setting isolates
artifact-level grounding (Section~\ref{sec:communication}) from file
organization:
producer-authored indices do not substitute for independently computed sidecar
fields.

In the configuration without dispatch, we replace independent worker dispatch with sequential execution of the
Modeler, Coder, and Writer in the Coordinator's session. Role
subworkspaces, stage order, quality checks, and \texttt{README.md}-based
handoffs remain unchanged, but task queues, asynchronous overlap, isolated
histories, and Coordinator-mediated forwarding are removed. This setting
therefore evaluates dispatch as the combination of delegation, context
isolation, and asynchronous scheduling rather than parallelism alone.

In the configuration without both metadata and dispatch, we apply both interventions simultaneously: all stages run sequentially in
the Coordinator's context, and the Writer receives no sidecar evidence. Role
subworkspaces and \texttt{README.md} indices are retained. This setting tests
whether metadata grounding and dispatch are complementary, or whether shared
context partially compensates for the missing sidecars.

\begin{table}[htbp]
\centering
\caption{Ablation results on MathModeling-Bench. All configurations use
DeepSeek V4.1 Flash.}
\label{tab:ablation}
\begin{tabular}{lcccc}
\toprule
\textbf{System} & \textbf{PSRF} & \textbf{MSQ} & \textbf{RVQ} &
\textbf{Final Score} \\
\midrule
Full system               & \textbf{72.57} & \textbf{64.33} & \textbf{80.46} & \textbf{70.87} \\
w/o metadata              & 67.03 & 61.16 & 64.89 & 64.65 \\
w/o dispatch              & 62.73 & 56.87 & 62.88 & 60.70 \\
w/o metadata and dispatch & 58.76 & 55.65 & 61.74 & 58.12 \\
\bottomrule
\end{tabular}
\end{table}

\subsection{Industry-Inspired Modeling Cases}
\label{sec:industrial-cases}

We examine \textcolor{blue}{2} industry-inspired tasks to make the demands of end-to-end
modeling concrete: sparse MIMO array design and open-world recognition of
encrypted traffic. Both use synthetic proxy experiments. The cases focus on
how mathematical choices affect computational checks and conclusions;
Appendix~\ref{app:industrial-cases} gives the formulations, experiment conditions,
and additional results.

\begin{figure}[!htbp]
\centering
\includegraphics[width=\linewidth]{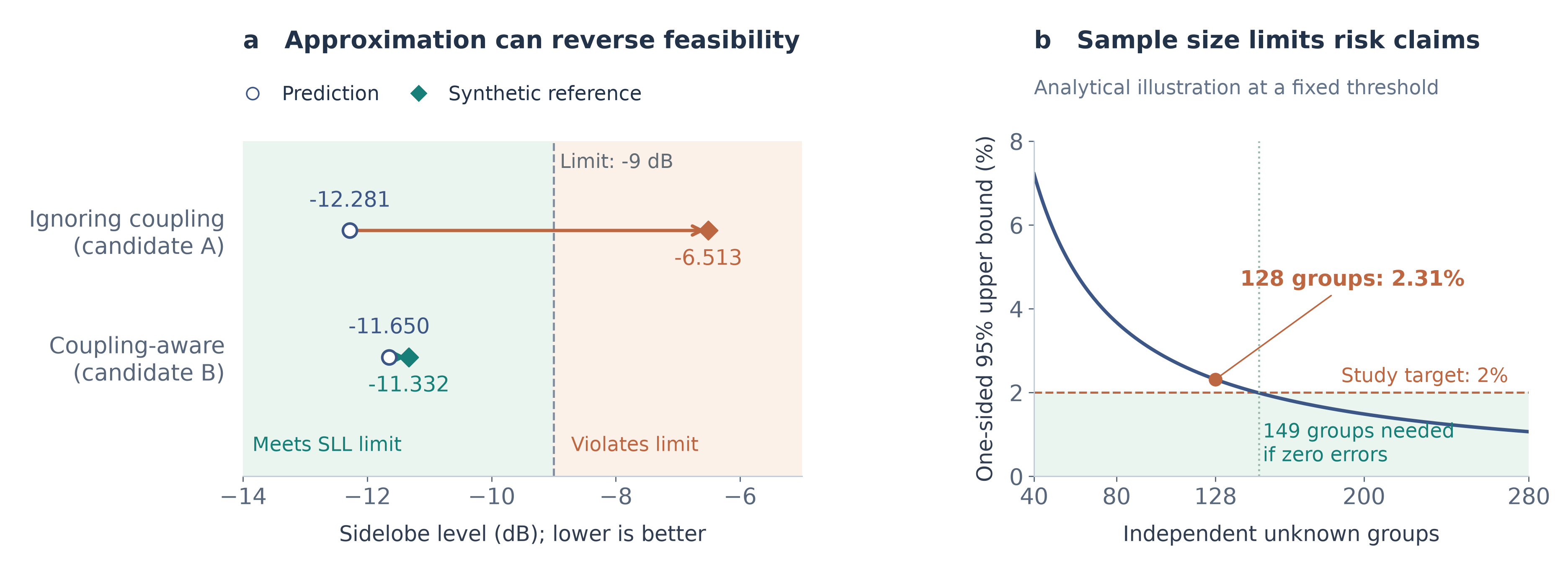}
\caption{Decisions that depend on the scope of computational evidence.
\textbf{(a)} Predicted and synthetic-reference SLL for \textcolor{blue}{2} different designs;
each arrow pairs evaluations of the same design.
\textbf{(b)} Analytical zero-error upper bound at a fixed threshold under
independent Bernoulli groups. The right panel is a sample-size calculation,
not a report of an observed zero-error test.}
\label{fig:industrial-decisions}
\end{figure}

In sparse-array modeling, the MIMO study combines reduced-order response models with coupling-aware
array optimization. A representative design obtained while ignoring coupling
has a predicted sidelobe level (SLL) of $-12.281$ dB, apparently satisfying
a $-9$ dB limit, but its synthetic-reference value is $-6.513$ dB.
A different coupling-aware design remains below the limit:
$-11.650$ dB predicted and $-11.332$ dB under reference reevaluation
(Figure~\ref{fig:industrial-decisions}a). In the broader search, 45 of
785 predicted Pareto candidates are classified as false-feasible.
Thus the modeling decision concerns both computational approximation and
the validity of the design constraints.

For encrypted-traffic recognition, on 118 synthetic validation groups over \textcolor{blue}{5} seeds, a statistical random forest
(RF) has macro-F1 $0.9139\pm0.0007$, compared with $0.9017\pm0.0173$ for
the dual-view DS-TCN. Combining RF labels with DS-TCN rejection scores improves
correct classification at the selected operating point but worsens overall
open-set ranking and near-unknown false acceptance, so the study retains RF.
Sample size also limits the claim that can be made: at a fixed threshold,
zero false acceptances in 128 independent unknown groups would still imply
a one-sided 95\% upper bound of 2.31\%, above the study's 2\% target
(Figure~\ref{fig:industrial-decisions}b).

These cases highlight a common requirement for MMA: assumptions and acceptance
criteria must remain explicit when model outputs pass into optimization,
selection, and reporting. They illustrate modeling dependencies and do not
provide an isolated test of the scheduler or a production deployment evaluation.

\section{Conclusion}
\label{sec:conclusion}

We presented ReasModel, an asynchronous multi-agent system that
coordinates specialized modeling, coding, and writing roles through explicit
artifact and metadata handoffs. We also introduced MathModeling-Bench, a new
Chinese-language, report-native benchmark that evaluates complete mathematical
modeling reports through problem-specific requirement fulfillment, general
modeling quality, and visual communication. Experiments on MathModeling-Bench
and MM-Bench demonstrate the effectiveness and efficiency of the proposed
workflow, while \textcolor{blue}{2} industry-inspired cases illustrate the importance of
preserving the physical, statistical, and implementation context of
computational results.


\section*{Acknowledgements}
We thank Bingyan Song of Xiangtan University for assistance with the preparation of the case studies.

\appendix
\section{Case Study}\label{app:exp-detail}

This appendix presents \textcolor{blue}{2} end-to-end case studies on CUMCM problems to
illustrate how \sysname{} coordinates formulation, implementation, validation,
and report generation. We highlight the key modeling decisions, failure-aware
revisions, evidence-gated results, and comparisons with strong contest solutions.


\subsection{Detailed Case Study: Solving the 2025 CUMCM Problem C}
\label{app:cumcm-2025c}

We conduct an end-to-end execution experiment targeting Problem C of the 2025 China Undergraduate Mathematical Contest in Modeling (CUMCM).

\subsubsection{Parallel formulation and execution}

The coordinator asynchronously dispatches the modeler, coder, and writer agents. The modeler translates the problem framework into executable specifications: Q1 constructs a dual-track model of continuous concentration trajectory-pass-probability trajectory; Q2 adopts a constrained one-dimensional optimal piecewise algorithm; Q3 builds an individualized risk surface; Q4 establishes a \textcolor{blue}{2}-stage multi-label screening framework of ``initial screening and subdivided determination''. Meanwhile, the coder sets up data schemas and metadata sidecar interfaces, and the writer completes the CUMCM report framework without filling pending computational results. Instead of forwarding full reasoning traces, the coordinator transmits structured digests containing adopted methods, required outputs, validation thresholds, and conclusion boundaries.

\begin{figure*}[!t]
\centering
\includegraphics[width=\textwidth]{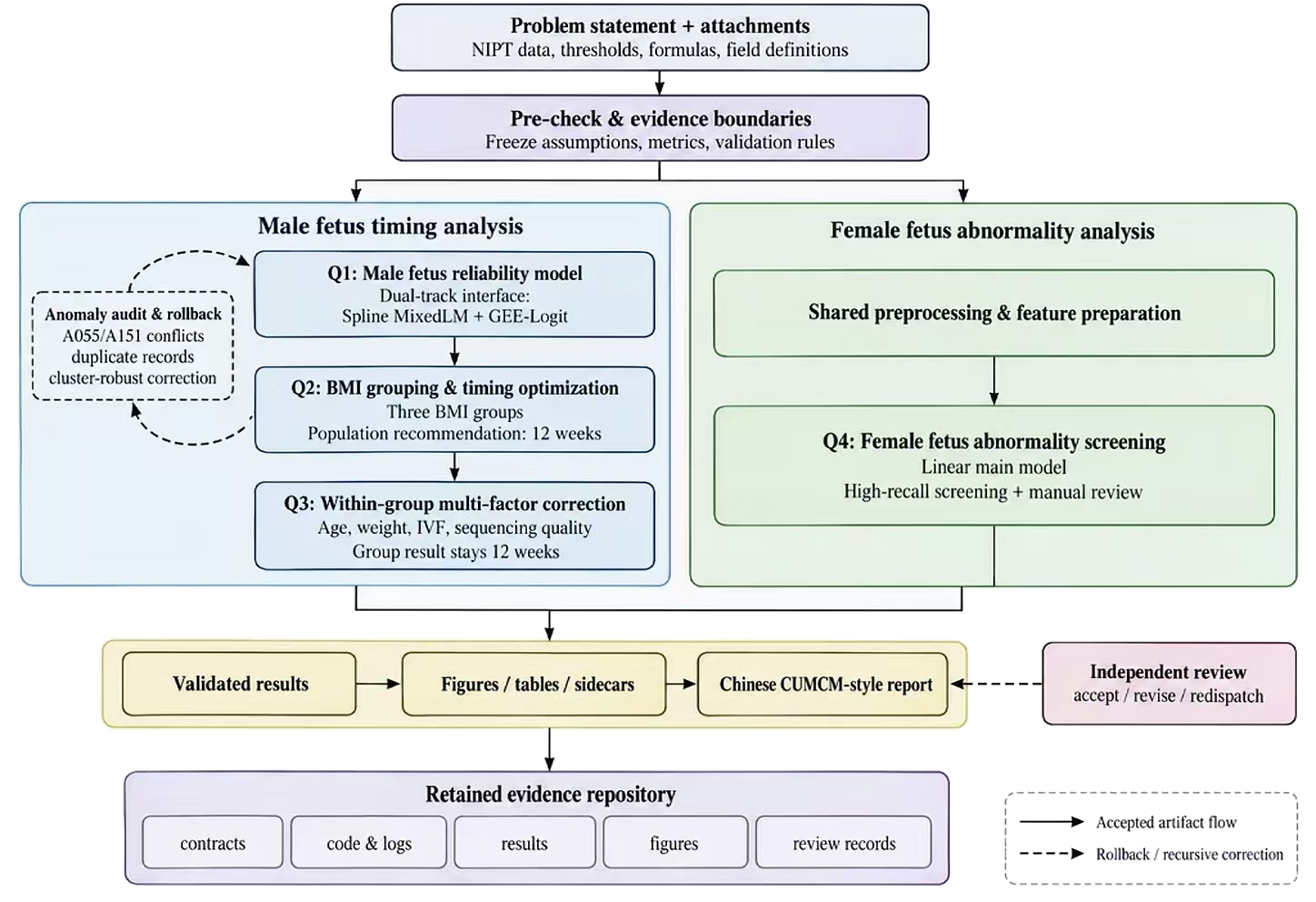}
\caption{Reconstructed evidence-gated workflow for CUMCM 2025 Problem C}
\label{fig:reconstructed-evidence-gated-workflow}
\end{figure*}

The overall execution logic of this parallel agent workflow is visualized in Figure~\ref{fig:reconstructed-evidence-gated-workflow}.

\FloatBarrier

\subsubsection{Failure-aware execution}

The first-round fitting for Q1 yields convergence warnings. The coder persists these cases into an anomaly log, without discarding observation samples or silently modifying decision rules. The log records \textcolor{blue}{2} categories of issues: time-consistency conflicts for subjects A055 and A151; duplicate detections with identical keys for 19 male-fetus groups and 15 female-fetus groups. The coordinator pauses downstream invocations over this feature and feeds fault summaries back to the modeler. The modeler adds \textcolor{blue}{3} fallback strategies: the original linear mixed-effects model, a sensitivity-analysis path with anomalous samples excluded, and robust standard-error estimation based on pregnant-subject clustering. After re-execution, 2 records follow the original rule, 2 pregnant subjects enter sensitivity-based exclusion, and the remaining 263 subjects adopt cluster-based robust correction. Only validated results are dispatched to downstream tasks. The writer documents this process as outcomes derived via algorithmic identification and sensitivity validation, rather than human-annotated ground truth.

\subsubsection{Evidence-gated results}

In Q1, the marginal effect of BMI is statistically significant under full-sample testing: for each 1-unit increase in BMI, the log-transformed value of Y concentration decreases by 0.0025 on average ($p<0.01$). Nevertheless, excluding samples A055 and A151 causes an effect-size shift exceeding 15\%. Accordingly, the conclusion is downgraded to statistically significant yet cautiously interpreted, and is not treated as a stable causal factor. In Q2, optimal time-points for \textcolor{blue}{3} subgroups all fall near week 12.0 under \textcolor{blue}{3} risk interfaces: continuous, generalized estimating equations (GEE), and consensus. \textcolor{blue}{4} out of \textcolor{blue}{5} weighting scenarios recommend week 12.0; one shifts to week 12.5 with no clinically significant inter-group difference. This conclusion passes robustness checks and is adopted as the primary result. In Q3, the optimal testing time-point for the high-risk subgroup of advanced-age, high-BMI subjects receiving in-vitro fertilization is postponed to week 13.5. Constrained by merely 9 samples within this subgroup, the report attaches a caveat: conclusions for this subgroup have limited evidence and require further clinical validation. Q4 subsequently employs tree-based models for \textcolor{blue}{5}-fold cross-validation across subgroups. The $F_2$-score for screening all abnormal samples reaches 0.623, whereas the $F_2$-score for compound chromosomal-abnormality samples drops to 0.41, failing to meet the preset performance threshold of 0.60. Hence compound-abnormality results are only presented as exploratory content in the appendix, and all model outputs are appended with manual-review flags.

\subsubsection{System-level implication}

This run is not a one-shot sequential pipeline across the modeler, coder, and writer agents. Where dependency constraints permit, modeling, code implementation, and report composition can advance in parallel; validation failures trigger targeted re-dispatch. The digest mechanism of the coordinator restricts inter-agent communication to confirmed conclusions. Role-specific artifacts are indexed via README files, and \texttt{.metadata} sidecars bind quantitative claims to generation scripts and upstream source files. Final auditing confirms: Q1 fully documents sensitivity fluctuation ranges for effect sizes; Q2's recommended time-points undergo multi-scenario robustness validation; Q3 adds limited-evidence warnings for small-sample-subgroup outcomes; Q4 retains manual-review constraints and explicitly clarifies model capability boundaries. This case therefore demonstrates that asynchronous role specialization combined with file-level provenance-tracking mechanisms enables end-to-end generation of complete, auditable modeling reports from heterogeneous contest inputs.

\subsubsection{Benchmarking against outstanding contest papers}

Compared with official national first-prize contest papers, the proposed ReasLab framework exhibits superior performance in core modeling dimensions, including longitudinal repeated data processing, dual-interface robustness verification, risk-sample feasibility tradeoff, population-individual heterogeneity differentiation, and standardized screening conclusion output.

The framework has \textcolor{blue}{3} key methodological advantages. First, the dual-track model adopted in Question 1 offers dual validation channels and sufficient modeling basis for subsequent analysis, which greatly alleviates single-model bias. Second, constrained by practical sample conditions, the results of Questions 2 and 3 eliminate invalid theoretical optimal solutions and avoid over-interpretation of feature effects, ensuring rigorous and credible conclusions. Third, the \textcolor{blue}{2}-stage multi-label screening framework with manual review mechanisms reasonably defines model application boundaries. It objectively reflects model performance and restricts the model to screening and early warning scenarios without overstating its diagnostic capacity.

Meanwhile, this method still has inherent limitations. Outstanding official papers show distinct strengths in global joint optimization, deep temporal censoring modeling, and ensemble model prediction performance. The detailed methodological and quantitative comparisons are presented in Table~\ref{tab:table-1-method-and-data-comparison-with-}.

\providecommand{\casemdcell}[2]{\textbf{Model:} #1\par \textbf{Data result:} #2}

\clearpage
\begin{table*}[!t]
\centering
\footnotesize
\setlength{\tabcolsep}{3.5pt}
\renewcommand{\arraystretch}{1.12}
\caption{Method and data comparison with official examples}
\label{tab:table-1-method-and-data-comparison-with-}
\begin{tabular}{@{}>{\raggedright\arraybackslash}p{0.08\textwidth}>{\raggedright\arraybackslash}p{0.29\textwidth}>{\raggedright\arraybackslash}p{0.29\textwidth}>{\raggedright\arraybackslash}p{0.28\textwidth}@{}}
\toprule
Problem & Official Example & ReasLab & Remarks \\
\midrule
Q1 & \casemdcell{Mixed-effects model (random intercept + slope);}{ICC ranges from 0.74 to 0.81; gestational age has a positive effect and BMI has a negative effect on Y-chromosome concentration.} & \casemdcell{B-spline linear mixed-GEE-Logit dual-track model;}{ICC = 0.7226; Y-chromosome concentration increases with gestational age and decreases with rising BMI.} & Both use mixed-effects models for longitudinal repeated measures, with consistent qualitative conclusions. Minor ICC differences from preprocessing are not directly comparable. \\[0.6em]
\midrule
Q2 & \casemdcell{Risk optimization using genetic algorithms or dynamic programming;}{Optimal gestational age of 11--12 weeks under lenient accuracy criteria.} & \casemdcell{Construction of a comprehensive risk function;}{Population optimal timing at 12 weeks.} & Both balance detection timing risks to solve optimal gestational age, with consistent qualitative results. Quantitative differences from risk functions and algorithms make outputs incomparable. \\[0.6em]
\midrule
Q3 & \casemdcell{Individualized correction implemented by introducing deep survival or extended mixed-effects models;}{BMI dominates between-group differences; age and weight only alter individual optimal timing.} & \casemdcell{Only within-group multi-factor individual correction under the original BMI grouping;}{Group-level recommendations remain at 12 weeks, mainly reflecting within-group individual differences.} & Both fix BMI grouping and correct within-group heterogeneity, with fully consistent group-level qualitative conclusions. Differences stem from individualized modeling implementations. \\[0.6em]
\midrule
Q4 & \casemdcell{Ensemble or corrected multi-label models with a review mechanism;}{Positioned as a screening tool; T21 is more difficult to identify than T18.} & \casemdcell{\textcolor{blue}{2}-stage multi-label screening model with a manual review mechanism;}{Model is only suitable for screening and review; T21 identification performance is weaker than T18.} & Both build multi-label screening models for imbalanced samples, with consistent chromosomal identification difficulty pattern. Performance metrics are incomparable due to different architectures. \\
\bottomrule
\end{tabular}
\end{table*}


\subsection{End-to-End Case Study: Solving the 2024 CUMCM Problem B}
\label{app:cumcm-2024b}

We examine an end-to-end run on Problem~B of the 2024 China Undergraduate Mathematical Contest in Modeling, which asks a manufacturer to couple incoming-lot sampling with component inspection, product inspection, replacement, disassembly, and reuse decisions. Q1 designs an anytime-valid test for a nominal defect rate of $p_0=0.10$; Q2 optimizes inspection and disassembly for a \textcolor{blue}{2}-component product; Q3 lifts the same accounting logic to an \textcolor{blue}{8}-component, \textcolor{blue}{3}-subassembly production tree; and Q4 propagates uncertainty in all stated defect rates through the Q2 and Q3 policy spaces. The Coordinator fixed \textcolor{blue}{3} boundaries before numerical execution: missing lot sizes were represented only through declared scenarios, inspection was treated as perfect because no error rates were supplied, and optimality claims were restricted to the explicitly enumerated policy class.

\subsubsection{Formulation, workflow, and failure correction}

The run separated mathematical formulation, executable verification, and report construction. The Modeler produced an anytime-valid Bernoulli evidence process for Q1 and a subtree state $(C,q,\sigma,D)$ for Q2--Q3, where $C$ is expected replacement cost, $q$ is pass probability, $\sigma$ records prior certification, and $D$ is the continuation credit available after disassembly. The Coder turned the accepted equations into fixed schemas, enumeration routines, invariance tests, posterior simulations, and paired metadata records. The Writer prepared the Chinese CUMCM report tree while quantitative prose remained gated on verified outputs.

The consequential failure occurred before the final enumeration: a one-level salvage formula valued recovered leaf components correctly in Q2 but omitted the descendants of an uncertified subassembly in Q3, whose own components may still be disassembled and reused after a failed reinspection. The derivation audit therefore replaced the shallow credit with a recursive continuation term; a constructed check valued the shallow and recursive credits at $5.0$ and $19.4$, showing that the omitted branch could materially alter policy values. Numerical execution stayed blocked until the revised recursion satisfied value bounds and $20{,}000$ randomized invariant checks, after which the implementation passed all $19$ entry-gate tests.

\subsubsection{Verified results and uncertainty bounds}

For Q1, the rejection and acceptance evidence thresholds were $20$ and $10$. Under a 500-item administrative cap, the acceptance, rejection, and unresolved probabilities were $0.9732$ at a true defect rate of $0.05$, $0.9986$ at $0.20$, and $0.8988$ at the boundary $p_0=0.10$, so the rule was described as anytime-valid at boundary crossing rather than globally optimal in average sample size. For Q2, exhaustive evaluation of $16$ policies in each of \textcolor{blue}{6} settings selected a single optimal action code per setting, with expected profits per qualified delivery ranging from $12.000$ to $21.679$. For Q3, all $27{,}648$ declared combinations were evaluated: the unique nominal optimum inspects all \textcolor{blue}{8} components, inspects and disassembles defective outputs at all \textcolor{blue}{3} subassemblies, skips final pre-market inspection, and disassembles returned defective final products, for an expected fulfillment cost of $139.778$ and an expected profit of $60.222$.

Q4 did not treat the tabled defect rates as known population parameters. Under the primary $N=100$ Jeffreys-prior scenario, the Bayes actions for all \textcolor{blue}{6} Q2 settings matched their plug-in actions and the Q3 Bayes action remained the nominal policy, although its posterior probability of being pointwise optimal was only about $0.424$ (expected value of perfect information $0.359$; fifth-percentile profit $54.336$). The final prose therefore distinguished ``best posterior mean action'' from ``most often optimal action,'' and treated effective sample size as a scenario rather than a recovered fact about the manufacturer.

\subsubsection{Audit trail and report outcome}

The retained artifacts show why the run was not a one-pass model--code--write chain: the derivation failure changed the state definition before computation, and later sensitivity and corner-stress results constrained how the Writer could describe stability. The numerical pipeline produced one data artifact and one metadata sidecar per analysis, then passed $13$ claim-integrity checks and a second-wave audit that downgraded \textcolor{blue}{5} overstrong claims. The resulting 44-page Chinese report passed independent review with no unresolved critical, major, or minor finding, showing how role specialization, failure redispatch, and file-level provenance turn a compact contest statement into an auditable chain from assumptions to executable decisions and final prose.

\subsubsection{Comparison with the Official Commentary and Displayed Solutions}
\label{app:cumcm-2024b-comparison}

The official commentary provides the numerical benchmark, while the \textcolor{blue}{3} displayed papers are treated as alternative accepted formulations. These sources were consulted only after the present run was completed. The comparison is summarized in Table~\ref{tab:cumcm2024b-official-method-comparison}.\footnote{Official problem commentary, ``Production-Process Decision Problems,'' China College Students Online, \url{https://dxs.moe.gov.cn/zx/a/hd_sxjm_sxjmstjp_2024sxjmstjp/241127/1980968.shtml}, accessed September 1, 2026.}\footnote{Official 2024 Problem B paper gallery, China College Students Online, \url{https://dxs.moe.gov.cn/zx/hd/sxjm/sxjmlw/2024qgdxssxjmjslwzs/2024btlw/}, accessed September 1, 2026.}

\providecommand{\cumcmcell}[2]{\textbf{Model:} #1\par \textbf{Data:} #2}

\begin{table}[!t]
\centering
\footnotesize
\setlength{\tabcolsep}{3.5pt}
\renewcommand{\arraystretch}{1.12}
\caption{Question-by-question comparison with the official example and the relative advantages provided by ReasLab.}
\label{tab:cumcm2024b-official-method-comparison}
\begin{tabular}{@{}>{\centering\arraybackslash}p{0.09\textwidth}>{\raggedright\arraybackslash}p{0.23\textwidth}>{\raggedright\arraybackslash}p{0.25\textwidth}>{\raggedright\arraybackslash}p{0.35\textwidth}@{}}
\toprule
Question & Official example & ReasLab & Relative advantage provided by ReasLab \\
\midrule
Q1 & \cumcmcell{binomial test and SPRT.}{fixed plan $(109,16)$; untruncated ASN $54$.} & \cumcmcell{mixture-likelihood e-process.}{thresholds $20/10$; cap $500$.} & ReasLab provides anytime-valid evidence under composite alternatives and an explicit \emph{unresolved} outcome, rather than forcing a binary decision at a fixed endpoint. \\[0.6em]
Q2 & \cumcmcell{enumeration of inspection and disassembly decisions.}{\textcolor{blue}{6} benchmark profits.} & \cumcmcell{recursive-renewal enumeration of $16$ policies per setting.}{profits match exactly in 3 of 6 settings.} & ReasLab provides renewal accounting that credits recovered components through the complete path to one qualified delivery, including their return to production. \\[0.6em]
Q3 & \cumcmcell{multistage inspection, disassembly, and return decisions.}{cost/profit $139.78/60.22$.} & \cumcmcell{recursive assembly-tree optimization over $27{,}648$ policies.}{cost/profit $139.778/60.222$.} & ReasLab provides the same selected policy and cost/profit after rounding, while certifying optimality by exhaustive enumeration rather than heuristic search. \\[0.6em]
Q4 & \cumcmcell{one-sided worst-case substitution.}{90\% confidence bound.} & \cumcmcell{Jeffreys-prior Beta posterior decision analysis.}{$N=50,100,200,500$; Q3 action probability $0.424$.} & ReasLab provides posterior-mean decisions, pointwise-optimality probabilities, lower-tail profit, and EVPI, giving a richer view of policy stability than a single worst-case bound. \\
\bottomrule
\end{tabular}
\end{table}
\FloatBarrier

\paragraph{Advantages and limitations.}
Relative to the official example and the displayed papers, ReasLab provides a more auditable end-to-end solution: recursive accounting preserves downstream recovery value, exhaustive enumeration verifies the stated policy classes, and posterior sensitivity quantifies how conclusions change under uncertainty. The main trade-offs are that the Q1 e-process is less familiar and requires explanation, the Bayesian Q4 analysis depends on declared sample-size and prior scenarios, and exhaustive Q2--Q3 enumeration becomes harder to scale for larger production networks. Within the dimensions evaluated here---recursive accounting, policy verification, uncertainty quantification, and reproducible evidence---the ReasLab solution is therefore relatively stronger, although the displayed papers remain more conventional and sometimes closer to the official accounting presentation.\footnote{Displayed papers B159, B195, and B196 are available from the official gallery cited above.}

\section{Industry-Inspired Modeling Case Studies}
\label{app:industrial-cases}

This appendix provides detailed formulations and synthetic proxy experiments
for the sparse MIMO array and encrypted-traffic case studies. It examines how
modeling assumptions, computational validation, and implementation fidelity
shape the conclusions that can be supported in each application.

\subsection{Case Scope and Modeling Dependencies}
\label{app:industry-scope}

\begin{figure}[!htbp]
\centering
\includegraphics[width=\linewidth]{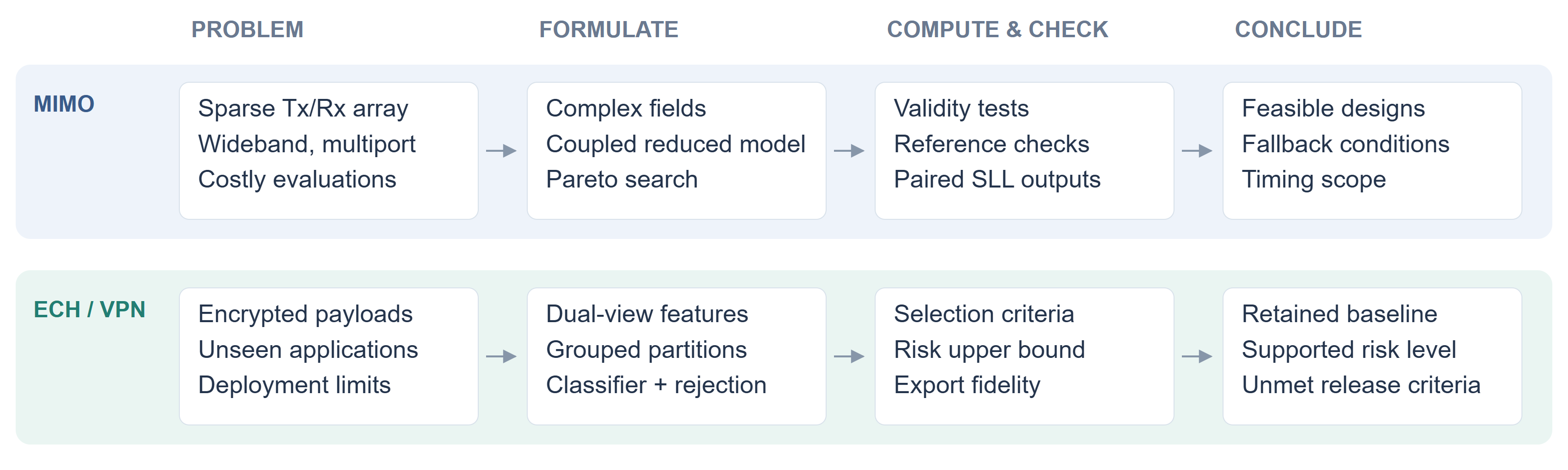}
\caption{Task dependencies in the \textcolor{blue}{2} industrial cases. The columns connect
the task specification to formulation, checks, and conclusions. Arrows show
logical dependencies; they do not encode execution time or individual role calls.}
\label{fig:industry-dependencies}
\end{figure}

The \textcolor{blue}{2} studies address different sources of difficulty in applied modeling.
Sparse-array design links approximation of a physical response to constrained
optimization; encrypted-traffic recognition links representation learning to
unknown-class rejection, finite-sample inference, and implementation requirements.
Figure~\ref{fig:industry-dependencies} summarizes these dependencies.
The diagram is reconstructed from the modeling studies and does not represent
a timestamped sequence of agent calls.

All numerical results in this appendix come from synthetic proxy experiments
reported for the \textcolor{blue}{2} projects. The MIMO study distinguishes a reduced-model
experiment (E1-A) from an array-optimization experiment (E1-B); their state,
port, and design counts should not be conflated. The ECH study distinguishes
validation, rejection calibration, and export-fidelity checks. These experiments
do not establish full-device electromagnetic accuracy, performance on operational
encrypted traffic, or successful deployment on a target device.

\begin{table}[!htbp]
\centering\small
\caption{Experimental settings relevant to the selected case evidence.}
\label{tab:industry-setting}
\setlength{\tabcolsep}{4pt}
\begin{tabular}{@{}p{.19\linewidth}p{.39\linewidth}p{.36\linewidth}@{}}
\toprule
Study & Experimental unit and scope & Main comparison \\
\midrule
MIMO E1-A & Synthetic reduced-model study; reported small-scale setting has
96 states and 6 ports. & Reduced response accuracy, validity checks, and fallback conditions. \\
MIMO E1-B & 42 candidate sites; 4 transmit and 4 receive elements; 360 unique
design evaluations per run. \textcolor{blue}{4} scenarios, \textcolor{blue}{3} search methods, \textcolor{blue}{5} seeds:
60 runs. & Random search, NSGA-II ignoring coupling, and coupling-aware NSGA-II. \\
ECH selection & 118 synthetic validation groups and \textcolor{blue}{5} fixed seeds. &
Statistical RF, statistical MLP, DS-TCN with statistics, and a conventional CNN. \\
ECH rejection & \textcolor{blue}{2} identity-disjoint evaluation folds with 60 and 68 unknown
groups; \textcolor{blue}{5} seeds. & RF labels with RF scores (R), versus RF labels with DS-TCN scores (H). \\
ECH export & Candidate implementations compared with their designated
same-seed reference. & Numerical and discrete decision fidelity before release criteria. \\
\bottomrule
\end{tabular}
\end{table}

\subsection{Sparse MIMO Arrays: Connecting Reduced Models to Feasible Designs}
\label{app:mimo}

The mathematical interface is defined as follows.
For a parameter configuration $\xi$, the response model uses
\begin{equation}
 A(\xi)X=B,\qquad S(\xi)=C_SX+D_S,\qquad E(\xi)=C_FX .
 \label{eq:industry-mimo-interface}
\end{equation}
Here $X$ contains the multiport state responses, $S$ is the complex scattering
response, and $E$ is the complex far field under the chosen excitation and
polarization conventions. Port order, phase, normalization, and the angular
domain must be preserved when the reduced response is used in array synthesis.
The normalized power pattern is a nonlinear postprocessing step,
\begin{equation}
 G_{\mathrm n}(\theta,\phi,f)
 =\frac{G(\theta,\phi,f)}{\max_{(\theta,\phi)\in\Omega}G(\theta,\phi,f)},
 \qquad
 G_{\mathrm n}^{\mathrm{dB}}
 =10\log_{10}\!\max(G_{\mathrm n},g_{\mathrm{floor}}).
\end{equation}
Composing magnitudes in place of complex fields can therefore alter interference
and the apparent satisfaction of beam constraints.

For model reduction, the study combines domain-decomposed characteristic basis functions, a weighted
POD construction, and overlapping local frequency models. Its weighted inner
product includes energy, port, and far-field contributions:
\begin{equation}
 Q=\frac{w_E M_e}{E_{\mathrm{ref}}^2}
  +w_S\widetilde C_S^H W_S\widetilde C_S
  +w_F\widetilde C_F^H W_\Omega\widetilde C_F,\qquad
 w_E+w_S+w_F=1.
 \label{eq:industry-weighted-pod}
\end{equation}
The retained state space requires $Q\succ0$; the output operators in this
expression are nondimensionalized. For a fixed candidate matrix $Y$, the
eigenproblem $Y^HQYu_i=\lambda_i u_i$ yields
$v_i=Yu_i/\sqrt{\lambda_i}$ for retained positive eigenvalues.
This construction targets the specified training inner product; it does not
guarantee a uniform bound on unseen far-field errors in dB.
Local models are checked using parameter-domain coverage, residuals, conditioning,
and applicable physical properties such as reciprocity and passivity.
Queries outside the supported conditions require a reference calculation or
model enrichment.

For array optimization, E1-B uses binary site selections and complex excitation weights. In the reported
configuration, the transmit and receive cardinalities are both \textcolor{blue}{4}.
The frozen predicted constraints require SLL below $-9$ dB, half-power beamwidth
at most $10^\circ$, and pointing error at most $5^\circ$.
Candidate designs are subsequently reevaluated with the synthetic reference.
The SLL pairs in Figure~\ref{fig:industrial-decisions}a refer to \textcolor{blue}{2} distinct
representative designs; they are not a before-and-after correction of one design.

Table~\ref{tab:mimo-fronts} reports scenario-level summaries across \textcolor{blue}{5} seeds.
The coupling-aware search has lower reported front migration and false-feasible
percentages in the nonzero-coupling scenarios. In the strong-coupling scenario,
the reported median false-feasible percentage changes from 75.0\% to 0.0\%.
A zero median does not mean that every run or every candidate is failure-free.
The edge-perturbation scenario retains a 28.6\% median for the coupling-aware
method, as visualized in Figure~\ref{fig:industry-extra}a.

\begin{table}[!htbp]
\centering\small
\caption{E1-B reference-reevaluation summaries reported across \textcolor{blue}{5} seeds.
Hypervolume is reported in dB\,$\cdot$\,degree. Front-migration values follow the study's normalization. These are scenario
summaries, not paired measurements of a single candidate.}
\label{tab:mimo-fronts}
\setlength{\tabcolsep}{4pt}
\begin{tabular}{@{}llrrr@{}}
\toprule
Scenario & Method & \shortstack{Reference\\hypervolume} & \shortstack{Median\\migration} & \shortstack{Median\\false-feasible} \\
\midrule
Weak coupling & Ignoring coupling & 1154.094 & 0.1054 & 16.7\% \\
Weak coupling & Coupling-aware & 1184.872 & 0.0172 & 0.0\% \\
Strong coupling & Ignoring coupling & 966.853 & 0.3810 & 75.0\% \\
Strong coupling & Coupling-aware & 1228.947 & 0.0358 & 0.0\% \\
Edge perturbation & Ignoring coupling & 1011.863 & 0.2603 & 66.7\% \\
Edge perturbation & Coupling-aware & 1117.267 & 0.0764 & 28.6\% \\
\bottomrule
\end{tabular}
\end{table}

Across all E1-B runs, the report retains 785 predicted Pareto candidates,
175 reference-feasible candidates, and 45 false-feasible candidates.
The 125 anomaly records include overlapping categories: 45 false-feasible,
79 large migrations, 15 reference multi-peak patterns, and 9 pointing violations.
One edge-perturbation run of the method ignoring coupling finds no
reference-feasible design; its 19 candidates remain in the record.
This evidence makes reference checking part of the modeling result and places
the favorable examples within the broader distribution of outcomes.

\subsection{Encrypted Traffic: Representation, Rejection, and Export}
\label{app:ech}

For its observable inputs, the ECH/VPN study uses packet direction, length, and inter-arrival sequences
together with 31 prefix-level statistics. Features are computed from packets
available at the decision time. Availability indicators distinguish a missing
measurement from a true zero, and derived statistics must be recomputed after
a permitted prefix perturbation. The study separates groups during data splitting
and unknown identities during rejection cross-fitting; overlapping prefixes
should not be treated as independent evidence about unknown-class risk.

Closed-set model selection compares \textcolor{blue}{4} candidates.
Table~\ref{tab:ech-models} shows the reported mean and seed standard deviation
for \textcolor{blue}{4} candidates. The selection rule requires DS-TCN to outperform the
strongest baseline on macro-F1 in at least \textcolor{blue}{4} of \textcolor{blue}{5} seeds, have a positive
mean difference, preserve worst-class recall within the specified tolerance,
and satisfy the stated paired comparison condition. DS-TCN wins on macro-F1
in only one seed. Its better NLL and calibration therefore motivate a separate
rejection experiment rather than a change of the primary classifier.

\begin{table}[!htbp]
\centering\small
\caption{Closed-set results on 118 synthetic validation groups. Entries
are mean $\pm$ standard deviation across \textcolor{blue}{5} seeds, not confidence intervals.
Higher macro-F1 and recall are preferable; lower NLL and ECE are preferable.}
\label{tab:ech-models}
\setlength{\tabcolsep}{4pt}
\begin{tabular}{@{}lcccc@{}}
\toprule
Model & Macro-F1 & Worst-class recall & NLL & ECE$_{15}$ \\
\midrule
Statistical RF & $0.9139\pm0.0007$ & $0.7500\pm0.0000$ & $0.4634\pm0.0340$ & $0.2330\pm0.0221$ \\
Statistical MLP & $0.9020\pm0.0136$ & $0.6125\pm0.0685$ & $0.6930\pm0.0750$ & $0.3726\pm0.0512$ \\
DS-TCN + statistics & $0.9017\pm0.0173$ & $0.6875\pm0.0442$ & $0.2937\pm0.0307$ & $0.0674\pm0.0176$ \\
Conventional CNN & $0.6062\pm0.0472$ & $0.2625\pm0.1355$ & $1.1379\pm0.0791$ & $0.2216\pm0.0262$ \\
\bottomrule
\end{tabular}
\end{table}

The open-world decision rule adds a rejection threshold to the RF predictions.
Let $\widehat y_g^{\mathrm{RF}}$ be the RF label for group $g$ and $r_g$
its known-class score. For a threshold $\eta$, the accepted label is
\begin{equation}
 h_\eta(g)=
 \begin{cases}
 \widehat y_g^{\mathrm{RF}}, & r_g\geq\eta,\\
 \bot, & r_g<\eta .
 \end{cases}
\end{equation}
For known groups $\mathcal K$ and unknown groups $\mathcal U$, the study uses
\begin{align}
 \widehat{\mathrm{CCR}}(\eta)
 &=\frac{1}{|\mathcal K|}
   \sum_{g\in\mathcal K}
   \mathbf{1}\{r_g\geq\eta,\widehat y_g^{\mathrm{RF}}=y_g\},\\
 \widehat{\mathrm{FBR}}(\eta)
 &=\frac{1}{|\mathcal U|}
   \sum_{g\in\mathcal U}\mathbf{1}\{r_g\geq\eta\}.
\end{align}
Thus rejecting a known target counts against CCR, while accepting an unknown
group contributes to FBR. OSCR is the area under the curve relating correctly
accepted known groups to unknown acceptance across thresholds.

\begin{figure}[!htbp]
\centering
\includegraphics[width=\linewidth]{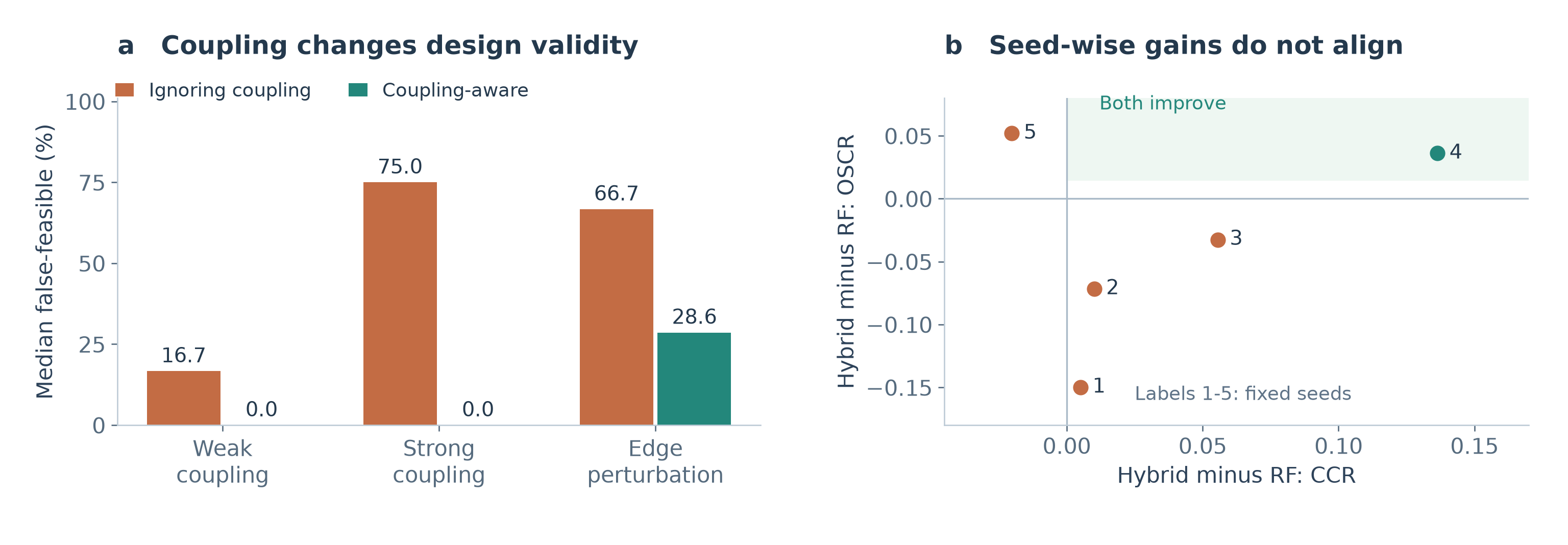}
\caption{Evidence beyond the representative examples.
\textbf{(a)} Reported median false-feasible percentages across \textcolor{blue}{5} runs in
each MIMO scenario. Zero denotes a median, not the absence of all failures.
\textbf{(b)} ECH hybrid-minus-RF differences for \textcolor{blue}{5} fixed seeds.
Improvement in one metric does not imply improvement in both.
The axes and points show reported differences, not newly estimated confidence intervals.}
\label{fig:industry-extra}
\end{figure}

Both rejection arms retain RF labels: R uses RF-derived scores and H uses
DS-TCN-derived scores. Across \textcolor{blue}{5} seeds, H improves mean CCR by 0.0374 but
reduces mean OSCR by 0.0332. Figure~\ref{fig:industry-extra}b shows the \textcolor{blue}{5}
paired differences: only one seed improves both quantities.
Near-unknown FBR also increases from 12.16\% to 15.14\%.
The reported joint selection criteria retain the RF-only backend.

Finite-sample support for the risk target is assessed as follows.
At a fixed threshold, with independent Bernoulli unknown groups and zero
false acceptances, the one-sided 95\% exact upper bound follows from
$(1-p_U)^n=0.05$:
\begin{equation}
 p_U=1-0.05^{1/n}.
 \label{eq:industry-zero-bound}
\end{equation}
For $n=60,68,128$, the bounds are 4.8703\%, 4.3099\%, and 2.3132\%,
respectively. An upper bound strictly below 2\% requires at least
149 independent zero-error groups. These are conditional sample-size
calculations, not observed zero-error results. Threshold selection on the
same data and pooled cross-fitted decisions require their own justification;
the pointwise formula alone does not certify an adaptively selected threshold.

Finally, the export checks test whether a compact implementation preserves the chosen
reference model. Table~\ref{tab:ech-export} separates exact reproduction, lossy
compression failures, and an unrun candidate. Relative to sklearn, the lossless
RF's maximum probability error is $5.551115\times10^{-17}$ with no label flips.
The zero errors in the first table row instead refer to the designated
lossless compact baseline. These reference choices should not be conflated.

\begin{table}[!htbp]
\centering\small
\caption{Worst export differences reported across \textcolor{blue}{5} seeds, relative
to the same-seed lossless compact baseline. Flip rates are fractions.
Q8 was not run after the upstream Q16 criterion failed.}
\label{tab:ech-export}
\setlength{\tabcolsep}{4pt}
\begin{tabular}{@{}lrrrrl@{}}
\toprule
Candidate & \shortstack{Max probability\\error} & \shortstack{Label\\flips} & \shortstack{Rejection\\flips} & \shortstack{Path\\flips} & Outcome \\
\midrule
Lossless RF & 0 & 0 & 0 & 0 & Pass \\
Tree prefix 250 & 0.230783 & 0.040186 & 0.023184 & 0 & Fail \\
Tree prefix 125 & 0.310874 & 0.095827 & 0.040186 & 0 & Fail \\
Q16 & 0.0279444 & 0.00309119 & 0.00154560 & 0.00275889 & Fail \\
Q8 & -- & -- & -- & -- & Not run \\
\bottomrule
\end{tabular}
\end{table}

Functional equivalence does not establish recognition quality or device
readiness. The overall study does not meet its release requirements,
despite the lossless export's successful numerical check.

\subsection{What These Cases Establish}

The MIMO case connects the fidelity of a mathematical approximation to the
feasibility of a proposed design. The ECH case connects feature construction,
model selection, rejection risk, and implementation fidelity.
Both make the assumptions behind a computational conclusion explicit.
They support application-level analysis of modeling decisions within the
reported synthetic settings. Attribution of individual decisions to specific
agent actions requires matching task inputs, generated files, human
interventions, and execution records; an isolated claim about scheduling
additionally requires a suitable system comparison.

\section{The \texttt{.metadata} Schema}\label{app:metadata_schema}

Every artifact produced by the Coder---a data file or a figure---has a sidecar that follows one of the \textcolor{blue}{2} templates below.

\begin{Verbatim}[fontsize=\small]
### Metadata for <output_filename>
- **Description**: <description>
- **Program**: `<program_name>`
- **Headers**: `<column_1>, <column_2>, ...`
- **Predecessors**: <input_file_1>
-- <input_file_2>
\end{Verbatim}

\begin{Verbatim}[fontsize=\small]
### Visual Metadata for <figure_filename>
- **Description**: <description>
- **Program**: `<program_name>`
- **Predecessors**: <input_file_1>
-- <input_file_2>
- **Autonomous Insights**:
- <metric_1>: <value_1>
- <metric_2>: <value_2>
\end{Verbatim}

The sidecar records what an artifact represents, how it was produced, which inputs it used, and the information needed to interpret it. It is generated together with the corresponding data or figure rather than written afterwards as a separate report. Its name is the full artifact name with \texttt{.metadata} appended, and the \textcolor{blue}{2} files are intended to be stored and read together. The figure variant additionally contains an \emph{Autonomous Insights} block.

The metadata-generation procedure combines task-level descriptions with information extracted or computed from the produced artifact and its underlying data. In particular, table headers are taken from the generated data, while figure statistics are computed from the data used to produce the figure. The writer reads the sidecar before referring to the corresponding artifact, and the coordinator checks reported quantitative claims against it (\S\ref{sec:communication}).

\subsection{Field-by-Field Specification}

The schema uses \textcolor{blue}{5} fields to describe artifact content, provenance, dependencies, and visual evidence.

The \emph{Description} field states what the artifact shows and its scope. For a table it gives the meaning of one record, the reported indicators with their units, and the time or region covered. For a figure it gives the objects being compared and the basis of the comparison. A label such as ``prediction results'' is too vague to be useful; it should let a reader who has not seen the analysis interpret the file.

The \emph{Program} field names the script that produced the artifact. It is recorded automatically so that the result can be traced and regenerated, and there is no need to inspect it.

For a table, the \emph{Headers} field lists the columns in the order in which they appear. The headers say which indicators are present, and the \emph{Description} says what they mean.

The \emph{Predecessors} field records the inputs that the artifact actually used: the raw data behind a processed table, the data behind a model result, or the result tables behind a figure. Each entry has a sidecar of its own, so the chain can be followed further upstream. Material that was considered but not used is left out.

For a figure, the \emph{Autonomous Insights} field reports the few statistics that support what the picture shows: how far \textcolor{blue}{2} schemes differ, when a curve peaks, how large and how variable a sample is, or whether an optimization result is feasible and optimal. The values come from the underlying data but are read back from the rendered image, so they describe the figure as drawn. Each value is given with its units and its comparison baseline.

\clearpage
\bibliographystyle{abbrv}
\bibliography{references}

\end{document}